\documentclass{article}

\usepackage{arxiv}

\usepackage[utf8]{inputenc} 
\usepackage[T1]{fontenc}    
\usepackage{hyperref}       
\usepackage{url}            
\usepackage{amsfonts}       
\usepackage{nicefrac}       
\usepackage{microtype}      
\usepackage{lipsum}
\usepackage{fancyhdr}       
\usepackage{graphicx}       
\usepackage{bm}
\usepackage{booktabs}

\usepackage[dvipsnames]{xcolor}
\usepackage[normalem]{ulem}

\usepackage{tikz}
	\usetikzlibrary{external, positioning, arrows.meta, calc, decorations.markings, shapes.geometric}
	\tikzexternaldisable{} 

\usepackage{pgfplots}
	\pgfplotsset{compat=newest}
	\usepgfplotslibrary{colorbrewer, colormaps, statistics, groupplots, fillbetween}
	\newlength{\figurewidth}
	\newlength{\figureheight}

\tikzset{%
    >={Stealth[length=2mm, width=1.75mm]}, 
    default line/.style={%
        semithick,
        line cap=round,
    },
    default dashed line/.style={%
        default line,
        dashed,
    },
    default marker line/.style={%
        default line,
        mark=*,
        mark size=0.75pt,
    },
}
\definecolor{Color1}{RGB}{31, 119, 180}
\definecolor{Color2}{RGB}{214, 39, 40}
\definecolor{Color3}{RGB}{0,0,0}

\usepackage{amsmath}
\usepackage{makecell}
\usepackage{cite}
\usepackage{xcolor}
\usepackage{placeins}
\usepackage[group-separator={,}]{siunitx}

\graphicspath{{media/}}     
\usepackage{amsmath}
\usepackage{xcolor}

\definecolor{ao(english)}{rgb}{0.0, 0.5, 0.0}

\title{Towards Spatio-Temporal Extrapolation of Phase-Field Simulations with Convolution-Only Neural Networks 
}

\author{Christophe Bonneville\thanks{Sandia National Laboratories, Livermore, CA 94550; \texttt{\{cpbonne,pmrobbe,hnnajm,csafta\}@sandia.gov}} 
\and 
\textbf{Nathan Bieberdorf}\thanks{Materials Sciences Division, Lawrence Berkeley National Laboratory \texttt{and} Department of Materials Science and Engineering, University of California, Berkeley, CA 94720, USA; \texttt{\{nbieberdorf,mdasta\}@berkeley.edu}} 
\and  \textbf{Pieterjan Robbe}\footnotemark[1]
\and  \textbf{Mark Asta}\footnotemark[2]
\and  \textbf{Habib N. Najm}\footnotemark[1]
\and  \textbf{Laurent Capolungo}\thanks{Los Alamos National Laboratory, Los Alamos, NM 87544; \texttt{laurent@lanl.gov}}
\and  \textbf{Cosmin Safta}\footnotemark[1]
}
\begin{document}
\maketitle

\begin{abstract}
Phase-field simulations of liquid metal dealloying (LMD) can capture complex microstructural evolutions but can be prohibitively expensive for large domains and long time horizons. In this paper, we introduce a fully convolutional, conditionally parameterized U-Net surrogate designed to extrapolate far beyond its training data in both space and time. The architecture integrates convolutional self-attention, physically informed padding, and a flood-fill corrector method to maintain accuracy under extreme extrapolation, while conditioning on simulation parameters allows for flexible time-step skipping and adaptation to varying alloy compositions. To remove the need for costly solver-based initialization, we couple the surrogate with a conditional diffusion model that generates synthetic, physically consistent initial conditions. We train our surrogate on simulations generated over small domain sizes and short time spans, but, by taking advantage of the convolutional nature of U-Nets, we are able to run and extrapolate surrogate simulations for longer time horizons than what would be achievable with classic numerical solvers. Across multiple alloy compositions, the framework is able to reproduce the LMD physics accurately. It predicts key quantities of interest and spatial statistics with relative errors typically below 5\% in the training regime and under 15\% during large-scale, long time-horizon extrapolations. Our framework can also deliver speed-ups of up to 36,000$\times$, bringing the time to run weeks-long simulations down to a few seconds. This work is a first stepping stone towards high-fidelity extrapolation in both space and time of phase-field simulation for LMD.
\end{abstract}


\section{Introduction}

Phase field modeling is a widely used computational technique for studying the evolution of microstructures in materials subjected to various external forces, such as chemical, mechanical, or electrical stimuli. By employing a diffuse-interface framework, it captures the behavior of phases, interfaces, and conserved quantities, with the system's dynamics governed by partial differential equations (PDEs) that describe continuum and interfacial processes \cite{LQChenReview,KarmaReview,SteinbachReview}. This approach avoids the complexities of explicitly tracking sharp interfaces but often demands fine spatial resolution to accurately model nano-scale phenomena, resulting in significant computational costs for large-scale simulations.

Simulating the phenomenon of de-alloying corrosion has served as a representative example for demonstrating some of the strengths and challenges associated with phase field modeling. The phase field method can naturally incorporate the thermokinetic processes of de-alloying (such as selective dissolution, long-range diffusion, surface transport, and capillary forces) to simulate complex morphologies and microstructures in qualitative agreement with experiments~\cite{geslin2015,Li:2022a,Lai:2022,bieberdorf2023,Lai:2022a,Bieberdorf2025}.  In doing so, phase field models have successfully revealed key mechanisms underlying the formation of de-alloying structures\cite{geslin2015,Lai:2022,bieberdorf2023,KerrBieberdorf2024,Bieberdorf2025}. However, such models have generally suffered from requiring extremely small grid spacing and time steps. Even when incorporated with advanced numerical techniques \cite{geslin2015,bieberdorf2023}, these models require substantial computational resources and are limited to smaller domains (e.g. a hundred $nm$ in size) and shorter time scales (e.g. a few $\mu$s) compared to experimental observations, necessitating extrapolation for validation.

The computational burden of phase field simulations becomes particularly problematic in scenarios requiring extensive direct numerical simulation, such as uncertainty quantification \cite{Ghanem:1991,Najm:2009a,LeMaitre:2010,10754/656260,Smith2013UncertaintyQ,Ghanem:2017}, inverse problems \cite{Kaipio:2005,Tarantola:2005,10754/656260,Galbally:2010,Fountoulakis:2016,peherstorfer2018survey}, and optimization tasks \cite{do1,do2,forrester2008engineering}. Surrogate models, including reduced-order models (ROMs), provide a more efficient alternative by reducing the dimensionality of the problem, often through techniques like proper orthogonal decomposition (POD) \cite{doi:10.1146/annurev.fl.25.010193.002543,rbm,benner2015survey,benner2017model,Safonov1988ASM}. Recently, machine learning methods, such as neural networks, have emerged as powerful tools for compressing high-fidelity data, offering superior performance in problems dominated by advection \cite{LEE2020108973,kutz_2017,BONNEVILLE2024116535,bonneville2024comprehensive}. 

A particular category of surrogate models for solving PDEs, namely neural operators~\cite{kovachki2021neural, li2021fourier, Lu_2021}, have recently gained popularity. These models establish mappings between function spaces through resolution-independent tensor-to-tensor transformations. Key frameworks include the Fourier Neural Operator (FNO) \cite{li2021fourier}, DeepONets \cite{Lu_2021}, Laplace Neural Operator (LNO) \cite{cao2023lno}, and Convolutional Neural Operator (CNO) \cite{raonić2023convolutional}. In dynamical systems, neural operators predict the state at future time steps based on the current state, often employing an iterative auto-regressive approach where the output feeds back as input for subsequent predictions. This method, which shares similarities with explicit time-stepping schemes, is particularly advantageous for handling non-parametric initial conditions, such as random fields, which traditional reduced-order models (ROMs) struggle to accommodate~\cite{Shih:2024,Jiang:2024a}. Neural operators have demonstrated their versatility across various applications, including fluid flow simulations \cite{li2021fourier, a16010024, fno_les}, electromagnetics \cite{Pornthisan:2024, Cai:2021}, fracture mechanics \cite{Goswami:2022}, meteorological forecasting \cite{pathak2022fourcastnet,10.1145/3592979.3593412}, phase field simulations \cite{oommen_deeponet,wen2022ufno}, and plasma physics \cite{gopakumar2023fourier}.

Neural operators have achieved impressive accuracy in previous studies, but recent advancements indicate that their performance can be further enhanced by integrating key components with U-Nets \cite{Ronneberger:2015u} and Vision Transformers (ViTs) \cite{dosovitskiy2021an}. Recent research has explored the use of U-Nets as feature extractors within an encoder-decoder framework, combined with neural operator bottlenecks \cite{wen2022ufno,diab2023udeeponet,10.1063/5.0158830}. Building on these developments, architectures such as ViTO \cite{ovadia2023vito} and DiTTO \cite{ovadia2023realtime} utilize ViTs to model interdependencies within compressed feature representations, where the compression is performed by a U-Net. More recently, \cite{bonneville_2025}, building on the work of \cite{Oommen:2024}, we proposed U-AFNOs for liquid-metal de-alloying (LMD) and corrosion simulations. This approach introduced a Vision Transformer (ViT) implemented in the Fourier space \cite{guibas2022adaptive} directly at the bottleneck of a U-Net, enabling higher computational efficiency. The results of that study demonstrated that this architecture significantly improves accuracy, allowing the model to reproduce the physical trends of LMD, even under highly chaotic conditions, while accelerating simulations by several orders of magnitude. The use of transformer-based architectures enhanced the model's ability to capture complex physical phenomena and reduced numerical instabilities.

Despite its success, this earlier work on LMD surrogates has several limitations. First, LMD simulations are computationally expensive and, in some cases, cannot be run to the desired time horizon due to prohibitively long runtimes, which can span months or even years on supercomputers. This limitation is particularly true for large domain size, which are desirable for fully understanding the LMD physics, but are typically unfeasible in practice due to the computational cost. Moreover, little research has explored the time extrapolation capabilities of LMD surrogates, particularly as the region of interest tends to grow beyond the fixed spatial domain size used during training. This poses a significant challenge for LMD, where de-alloying expands dynamically over time and may progress into the material with sometimes limited barriers. Additionally, the model proposed in \cite{bonneville_2025} is constrained by the time step sizes and material properties present in the training data, limiting its flexibility to handle a broader range of input conditions. Another drawback is that the U-AFNO model requires initialization with initial conditions generated by running the numerical solver for a short period. While the U-AFNO itself is computationally efficient, this initialization step is time-consuming and reduces the effective speed-up achieved by the surrogate simulation.

To address the limitations of previous models, particularly the U-AFNO introduced by \cite{bonneville_2025}, we propose several key modifications. First, we remove all fully connected layers typically present in the Vision Transformer bottleneck, retaining only the self-attention layers \cite{vaswani2017attention,dosovitskiy2021an,bahdanau2015neural}, which weigh the importance of different regions within the field. These attention layers are implemented exclusively using convolutional kernels. Additionally, we incorporate parametric conditioning, similar to the approach introduced by \cite{Oommen:2024}, enabling the model to handle a variety of simulation parameters. In particular, we develop the model to be sensitive to the alloy's initial chemistry, which is known to have a large impact on the rate and topology of the dealloying front \cite{geslin2015,Lai:2022,bieberdorf2023,Bieberdorf2025}.  This conditioning scales only the channel dimensions in the convolutional layers of the U-Net, ensuring that the entire architecture is fully convolution-based. As a result, our proposed model becomes independent of the spatial domain size, as convolution kernels can operate across fields of any dimensions (although the resolution is assumed to remain fixed).

To further enhance the model's capabilities, we introduce circular padding to enforce physical boundary conditions. We also implement a corrector method designed to maintain accurate species concentrations in regions that have not yet undergone corrosion, improving stability in long-term auto-regressive surrogate simulations. These modifications allow the model to be applied to domains of any size while maintaining high stability. Consequently, the model can extrapolate and predict dealloying patterns for much larger domains and longer time horizons than those present in the training data. Notably, our proposed model achieves accurate predictions for time horizons that would be impractical to include in the training data due to the excessive computational cost of generating such simulations.

To further improve efficiency, we also introduce a generative AI method~\cite{Kingma:2022,Rezende:2016,NIPS2014_5ca3e9b1,Ho:2020} for producing synthetic initial conditions, eliminating the need for numerical solver-generated initial conditions as required in \cite{bonneville_2025} and \cite{Oommen:2024}. This approach significantly accelerates the surrogate simulation process. Specifically, we reuse the U-Net architecture from the proposed surrogate and integrate it into a diffusion model~\cite{Ho:2020}. Diffusion models are capable of generating synthetic yet realistic solution fields based on conditional parameters (e.g., prompts). In this paper, we propose to use diffusion models to generate field images based on input chemical parameters, which can then serve as initial conditions for the surrogate model. The main contributions of this paper are therefore the following:
\begin{itemize}
    \item A novel convolution-only conditional U-Net model that is specifically tailored to handle physical constraints and model inputs, and remains stable while extrapolating both in time and space, allowing simulation time-horizons much beyond what is achievable with both direct numerical solvers and existing surrogate methods.
    \item A novel approach to generate suitable initial conditions using diffusion models, allowing for completely bypassing the need for solver-generated initial conditions, therefore dramatically increasing the effective speed-up of a surrogate simulation.
\end{itemize}

The paper is organized as follows: Section \ref{pf_descrip} introduces the phase-field model for LMD considered in this paper, and also briefly discusses the numerical solver bottleneck. Section \ref{surrogate} describes the surrogate model (subsections \ref{unet-arch}, \ref{flood_fill} and \ref{initial_cond} detail the U-Net architecture, the field corrector approach, and the initial condition generation, respectively). Section \ref{trainingtesting} discusses how the model is trained and tested, as well as evaluation metrics. Finally, section \ref{results} presents the model predictions and performance. 

\section{Phase Field Model Description}
\label{pf_descrip}
This study employs a high-fidelity phase field model to simulate de-alloying of a binary alloy composed of species \(A\) and \(B\), exposed to a corrosive liquid bath of pure species \(C\). The model and parameters considered here follow those proposed in the liquid metal dealloying models from \cite{geslin2015,bieberdorf2023}, where species \(A\), \(B\), and \(C\) represent tantalum, titanium, and copper, respectively. The solid and liquid phases are tracked using a non-conserved phase field variable \(\phi\), where \(\phi = 1\) represents the solid phase, \(\phi = 0\) represents the liquid phase, and \(0 < \phi < 1\) denotes the diffuse interface between the two phases. The mole fractions of species \(A\), \(B\), and \(C\) are represented by conserved phase field variables \(c_\text{A}\), \(c_\text{B}\), and \(c_\text{C}\), respectively, constrained by \(c_\text{A} + c_\text{B} + c_\text{C} = 1\) everywhere in the domain.

The evolution of the phase field is governed by \cite{SteinbachReview, Steinbach1999}:
\begin{equation}
\frac{\partial \phi}{\partial t} = -\tilde{M}_\phi \frac{\pi^2}{8 \eta} \frac{\delta F}{\delta \phi}
\end{equation}
where \(\tilde{M}_\phi\) is the interface mobility, \(\eta\) is the diffuse interface width, and \(\delta F / \delta \phi\) is the functional derivative of the free energy functional \(F\). The mole fractions of species \(A\) and \(B\) evolve according to a continuity equation:
\begin{equation}
\frac{\partial c_i}{\partial t} = \nabla \cdot \sum_{j=\text{A,B}} M_{ij}(\phi) \nabla \left( \frac{\delta F}{\delta c_j} \right)
\end{equation}
where \(M_{ij}(\phi)\) is the phase-dependent solute mobility, and \(\delta F / \delta c_j\) represents the diffusion potential of species \(j\). Species \(C\) is determined via \(c_\text{C} = 1 - c_\text{A} - c_\text{B}\).

The free energy functional \(F\) is defined as:
\begin{equation}
F = \int_\Omega \left( f_\text{phase}(\phi) + f_\text{chem}(\phi, c_\text{A}, c_\text{B}) \right) \, \text{d}V
\end{equation}
where \(f_\text{phase}\) describes the diffuse interface energy, and \(f_\text{chem}\) accounts for the chemical energy density.

The computational domain is a 2D square (\(\Omega = [0, 102.4] \times [0, 102.4]\) nm$^2$) discretized into a \(512 \times 512\) grid with spacing \(\Delta x = 0.2\) nm. The binary alloy is initialized in the lower portion of the domain with an ideal spatially-uniform composition $(c_\text{A}, c_\text{B}=1-c_A)$. In this study, we consider $c_A\in[0.2, 0.4]$, and correspondingly, $c_B\in[0.6, 0.8]$. The species fields are perturbed by low-amplitude random noise, while the liquid phase in the upper portion of the domain is pure \(c_\text{C} = 1\). Dirichlet boundary conditions (\(\phi = 0\), \(c_\text{C} = 1\)) are applied at the top edge, Neumann boundary conditions (zero normal gradient) at the bottom edge, and periodic boundary conditions on the left and right edges.

In the direct numerical solver, the non-conserved variable (\(\phi\)) is updated using forward Euler time integration, while conserved species (\(c_\text{A}\), \(c_\text{B}\)) are updated using a semi-implicit spectral method to address numerical stiffness. The stiffness arises from the fine spatial resolution of the numerical mesh and the fourth-order derivative terms in the species evolution equation, which are introduced by the Laplacian in the chemical energy density. To ensure numerical stability and accuracy, the simulations employ a very small time step size of \(\Delta t = 10^{-12}\text{s}\). Thermodynamic and kinetic parameters are consistent with those used in \cite{geslin2015, bieberdorf2023, bonneville_2025}. For further information on the phase-field model and the numerical scheme briefly summarized in this section, we refer the reader to~\cite{bieberdorf2023,KerrBieberdorf2024}.

\section{Surrogate Model}
\label{surrogate}

\subsection{U-Net Architecture}
\label{unet-arch}
U-Nets~\cite{Ronneberger:2015u} are powerful tensor-to-tensor regression architectures widely employed in various deep learning applications~\cite{Jin:2017,Roth:2015,Guo:2016,Bhatnagar:2019}, originally introduced for biomedical image segmentation tasks. They are specialized types of residual convolutional neural networks (CNNs)~\cite{he2016deep}, characterized by their distinct U-shaped encoder-decoder structures interconnected by skip (residual) connections~\cite{drozdzal2016importance}. These skip connections directly link layers from the encoders to corresponding layers in the decoders, enabling efficient propagation of gradient signals during training, mitigating the vanishing gradient problem, and enhancing convergence speed and accuracy. Recently, U-Nets have been increasingly adopted in physical modeling and simulations governed by partial differential equations (PDEs), especially for autoregressive prediction tasks~\cite{Thuerey:2020,Stachenfeld:2022}. In these settings, U-Nets iteratively map spatial input fields to subsequent simulation states, demonstrating a remarkable capability to accurately capture complex spatio-temporal dynamics inherent to physical systems.

\begin{figure}[!ht]
\centering
    \includegraphics[width=1\textwidth]{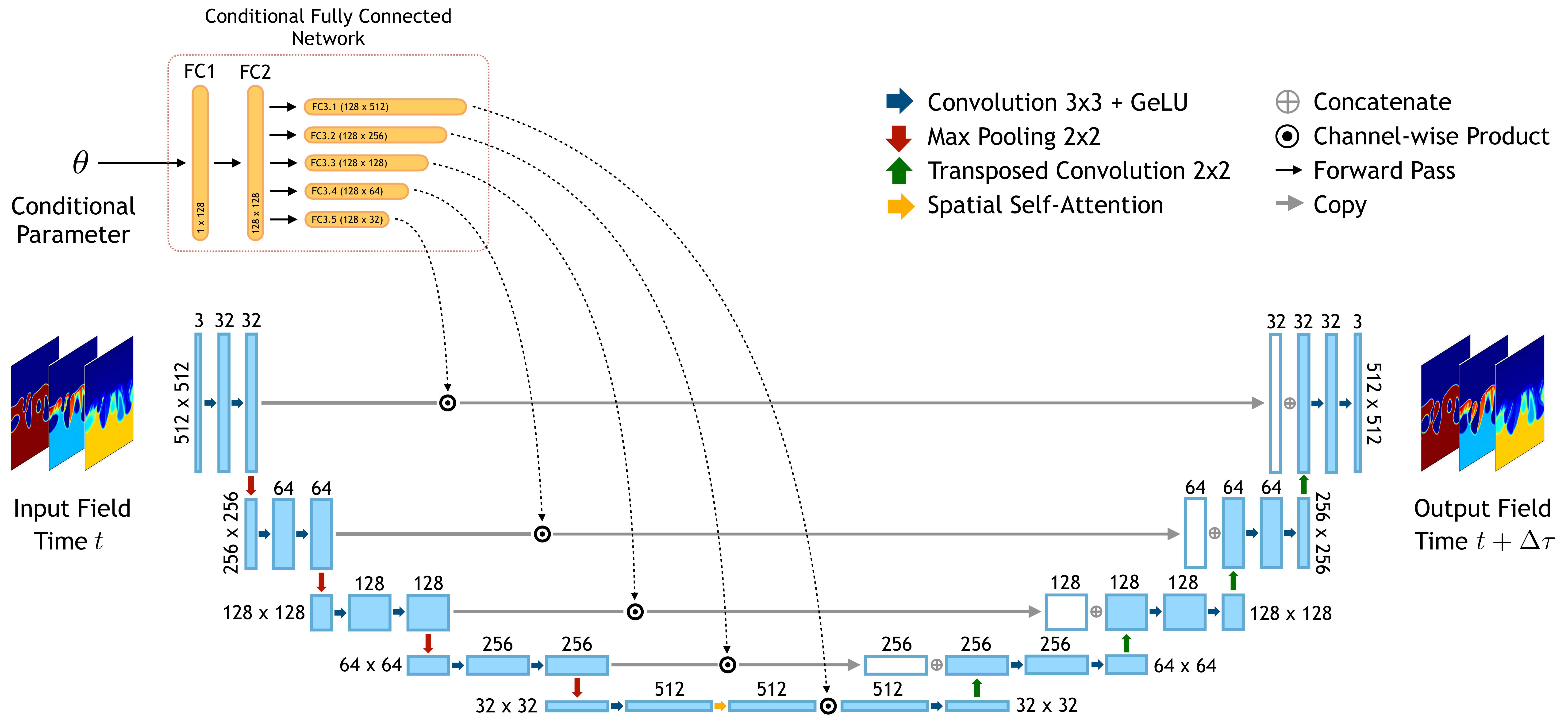}
    \caption{U-Net Architecture - The model takes as input the field at time $t$ and outputs the field at a future time step $t+\Delta\tau$, where $\Delta\tau$ is a conditional parameter included in $\theta$. The conditional network outputs vectors that scale each channels of the residual connections during reconstruction. The U-Net color scheme is analogous to the one employed in the original U-Net paper~\cite{Ronneberger:2015u}. }
    \label{fig:cunet}
\end{figure}

Note that U-Nets are not technically neural operators (such as Fourier Neural Operators (FNO)~\cite{li2021fourier} or Deep Operator Networks (DeepONet)~\cite{lu2021deeponet}) and therefore do not inherently possess resolution-invariant properties. Consequently, they typically must be retrained when changing grid resolutions. However, precisely because U-Nets are purely convolution-based architectures, they can be trained on snapshots of physical fields taken over relatively small spatial domains and subsequently applied to larger domains, provided that the  resolution remains consistent. This property emerges directly from the translational invariance and locality of convolutional filters~\cite{long2015fully, goodfellow2016deep}. The U-Net architecture employed in this paper is detailed in Figure~\ref{fig:cunet}. Since the phase field and the species fields are bounded into the $[0,1]$ interval, we wrap the output of the U-Net in a sigmoid function (as done in our previous work \cite{bonneville_2025}). This strictly guarantees that no field predicted by the surrogate can exhibit values outside this interval.

\subsubsection{Parametric Conditioning}

In the present work, we employ a conditional U-Net architecture, explicitly conditioned on parameters such as the prediction time step size and material properties. In principle, this conditioning framework can be generalized to incorporate additional physical or numerical parameters. This conditional formulation allows our architecture to predict solutions at arbitrary future times, effectively skipping multiple intermediate simulation steps, and to seamlessly conduct surrogate simulations under varying material conditions without retraining. This conditioning approach is directly inspired by recent advances such as diffusion models, where time-conditioned U-Nets are employed for generative tasks~\cite{ho2020denoising, nichol2021improved}, the DiTTO framework~\cite{ovadia2023realtime}, and the time-conditioned U-Net for material modeling proposed by Oommen et al.~\cite{Oommen:2024}. Conditioning is implemented using a fully connected neural network that takes as input a conditioning vector $\theta$. This vector can be of any dimension and introduces multiple parameters, but in this paper, we restrict $\theta$ to contain the time step size $\Delta\tau$ skipped at each U-Net forward pass ($\Delta\tau\in[5\cdot10^4\Delta t, 10^5\Delta t]$), and the initial species concentration $c_A\in[0.2, 0.4]$ (i.e. the $c_A$ concentration when no corrosion is unfolding). This fully connected network consists of two hidden layers, each comprising 128 neurons with SiLU activation functions. The network's output is then passed through five distinct linear layers, producing scaling vectors of sizes 32, 64, 128, 256, and 512, respectively, consistent with the number of feature channels at different convolutional layers in the U-Net. These scaling vectors modulate the outputs of each convolutional layer within the U-Net prior to concatenation via skip connections, providing a powerful and flexible mechanism to incorporate parametric dependencies directly into the network's predictions.

\subsubsection{Attention Layers}

We integrate attention mechanisms into the bottleneck convolutional layer of the U-Net architecture. Attention mechanisms, initially introduced for natural language processing tasks~\cite{bahdanau2015neural,vaswani2017attention}, have recently been successfully adapted to computer vision applications~\cite{wang2018nonlocal,dosovitskiy2021image}. These mechanisms can enhance model performance by allowing neural networks to dynamically focus on the most informative features within images or feature maps. Intuitively, attention layers enable the network to selectively emphasize relevant channels or spatial regions, thus improving the representation of important details and enabling better generalization. The incorporation of attention in our architecture is inspired by recent work, including diffusion models~\cite{ho2020denoising,nichol2021improved} and the DiTTO framework~\cite{ovadia2023realtime}, both of which have demonstrated the effectiveness of attention for enhancing predictions of complex spatio-temporal phenomena. In this paper, we implement the following spatial self-attention in the U-Net bottleneck:


\textbf{Self attention}: Captures long-range dependencies across spatial locations. Following the self-attention approach of~\cite{zhang2019self}, given the feature map $X\in\mathbb{R}^{C\times H\times W}$ (where C is the number of channels, here 3, and H and W are the field spatial dimensions), we first project it into query, key, and value tensors using three convolutions with $1\times1$ kernels:
\begin{equation}
Q=\text{Conv}_Q(X)\quad K=\text{Conv}_K(X), \quad V=\text{Conv}_V(X)
\end{equation}
with $Q,K\in\mathbb{R}^{C'\times H\times W}$ (in this paper, $C'=C/8$), and $V\in\mathbb{R}^{C\times H\times W}$. Reshaped into sequences, the spatial attention map is computed as:
\begin{equation}
A_s=\text{softmax}(Q^\top K)\in\mathbb{R}^{(HW)\times(HW)}
\end{equation}
The attention-weighted output is then obtained by:
\begin{equation}
X'= \gamma VA_s^\top+X
\end{equation}
where $\gamma$ is a learnable scalar initialized as zero, providing residual learning. Due to the higher computational complexity associated with spatial attention, we limit its use exclusively to the most down-sampled layer of the U-Net (i.e. the bottleneck). Note that all attention computations in our implementation rely solely on convolutional operations, which effectively preserve the model's flexibility and ensure input-size independence. We have shown in our earlier work \cite{bonneville_2025} that stacking attention layers through a Vision Transformer (ViT) within a U-Net bottleneck can considerably improve phase-field simulation performance. However, a ViT has fixed input sizes due to its reliance on fully connected layers with fixed dimensions. By keeping only the attention layers here, the U-Net remains flexible to various input sizes, while retaining the strength of ViTs \cite{app15074019,zhang2024,ZHUANG2025117623,Schlemper:2019}. 

\subsubsection{Padding Strategy}
\label{padding}

Due to the convolutional nature of the U-Net architecture described previously, padding operations are required throughout the network to ensure consistency between input and output spatial dimensions. Each convolutional layer employs a kernel size of $3\times3$, thereby requiring padding of width 1 along the left, right, top, and bottom edges of the input tensor so that the feature maps maintain constant height and width. The most common and widely employed padding strategy with convolutional neural networks is zero-padding, \textit{i.e.} adding rows and columns of zeros around the edges of the tensor~\cite{Ronneberger:2015u,goodfellow2016deep}. In the present work however, we do have knowledge about the physical constraints governing the boundaries, and thus zero-padding may not be the optimal choice. In particular, employing a customized padding strategy tailored to the expected boundary conditions of our fields is desirable~\cite{Alguacil:2021,Ren_2022}. 

\begin{figure}[!ht]
\centering
    \includegraphics[width=0.75\textwidth]{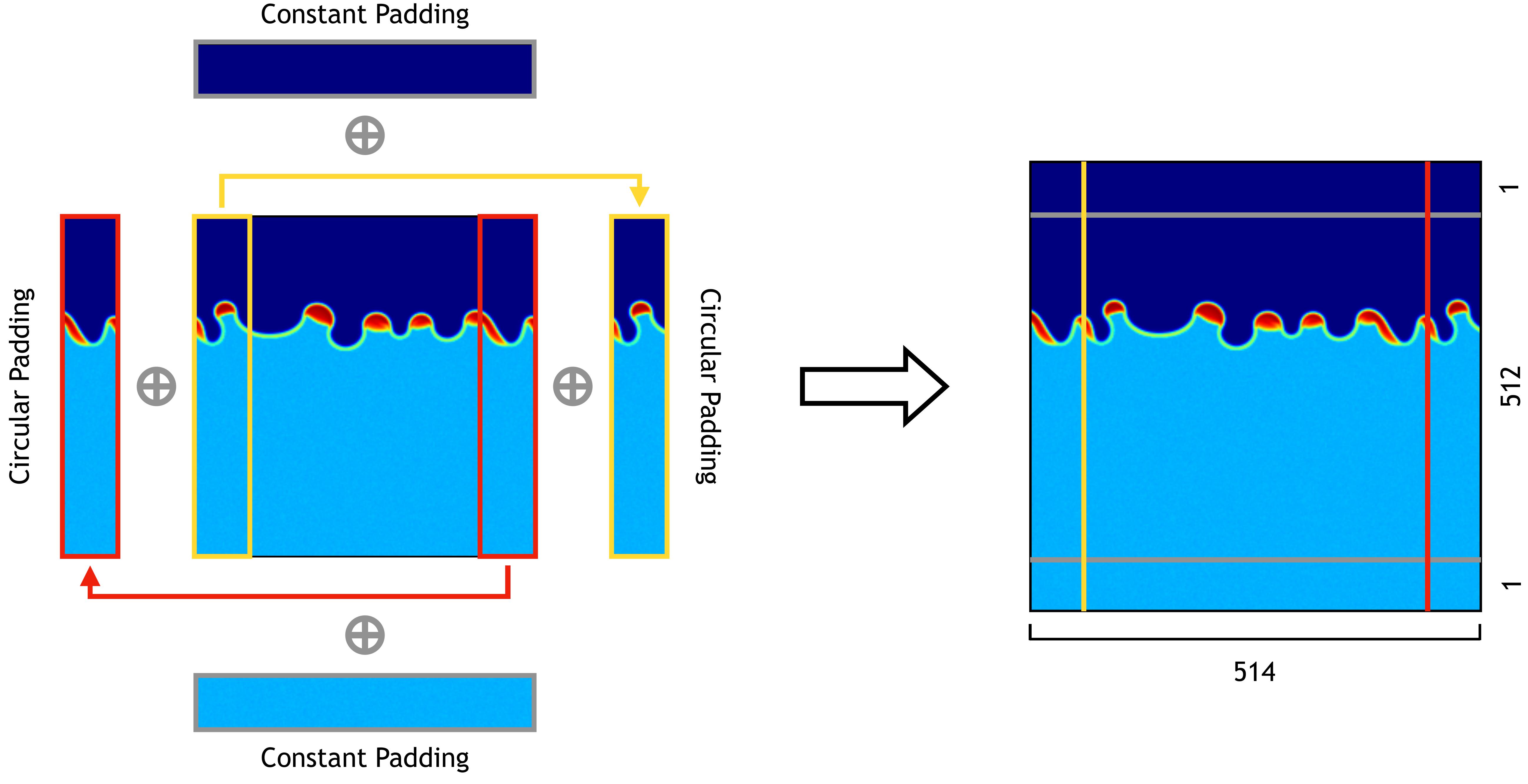}
    \caption{Padding in each U-Net convolutional layers. The width of the padded tensors is exaggerated for illustration purposes. In practice, since the convolution kernels in the U-Net are $3\times3$, the padding width required is only 1.}
    \label{fig:circ_padding}
\end{figure}

We employ \textit{circular padding} along the horizontal (left and right) edges, a padding method that wraps the input tensor around itself by replicating values from the opposite edges. Concretely, this means that values from the right edge of the tensor are appended to the left side and vice versa, effectively modeling periodic boundary conditions. Additionally, we apply zero-padding at the top boundary, as both the phase field and the species concentration are expected to remain zero (or nearly zero) there. At the bottom boundary, we replicate (pad with) the values present at the boundary itself, reflecting our expectation that the metal composition remains constant and identical just below the computational domain. This carefully chosen padding scheme explicitly incorporates known physical behaviors into our neural architecture,  improving model accuracy and generalization capabilities. A depiction of circular padding is shown in Figure~\ref{fig:circ_padding}.

\subsection{Flood-Fill Approach for Cleaning the Field Snapshots}
\label{flood_fill}

During time integration with the numerical solver, species concentration values in uncorroded spatial regions (i.e. regions sufficiently distant from the evolving liquid-metal interface) remain unchanged. Consequently, these regions are expected to exactly retain their original concentration values as the simulation is progressing. However, since the U-Net surrogate is a data-driven model, it may not perfectly replicate the identity mapping in these uncorroded regions, potentially introducing small deviations or noisy and un-physical artefacts between the input and output fields. To address this issue, we employ a flood-fill algorithm to systematically identify the uncorroded region after each forward pass of the U-Net and replace the predicted concentration values in these regions with the known, ideal concentration values. This correction ensures that the surrogate model adheres to the physical constraint that no changes should occur in uncorroded areas, thereby improving the consistency and reliability of the predictions.

\begin{figure}[!ht]
\centering
    \includegraphics[width=1.\textwidth]{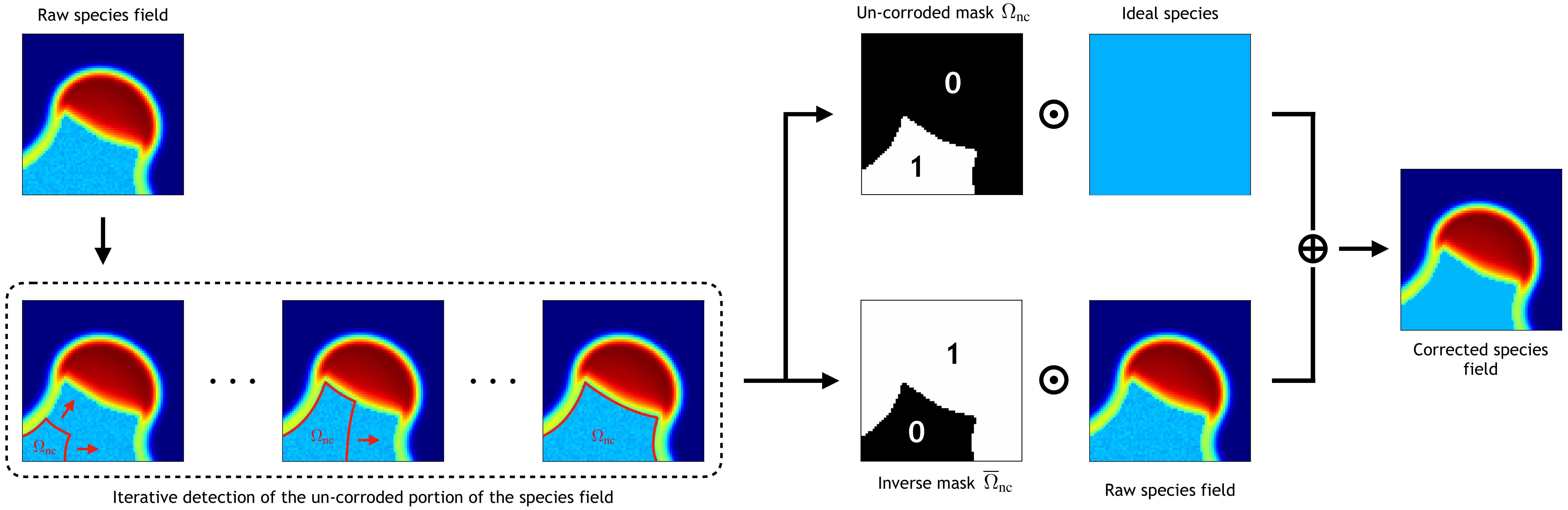}
    \caption{Flood-fill algorithm for field cleaning. The approach first identifies where the un-corroded portions of the species fields stop. It subsequently generates corresponding masks to replace the un-corroded (yet potentially corrupted) nodes with the known ideal species value.}
    \label{fig:masking}
\end{figure}

The flood-fill algorithm~\cite{levoy1981area,heckbert1990seed} is used to identify the uncorroded region, denoted as \(\Omega_{\text{uc}}\). We initialize the flood-fill process with \(\Omega_{\text{uc}}^{(0)}\), consisting solely of the spatial point at coordinates \((x=0.5, y=0)\) (i.e. the center of the bottom boundary). At each iteration, the uncorroded region is expanded by including all neighboring spatial points adjacent to the current region \(\Omega_{\text{uc}}^{(k)}\) that satisfy the proximity criterion to the ideal species concentration value. Formally, the iterative update is expressed as:
\[
\Omega_{\text{uc}}^{(k+1)} = \Omega_{\text{uc}}^{(k)} \cup \left\{(x,y) \in \partial \Omega_{\text{uc}}^{(k)} \, \text{s.t.} \, |c(x,y) - c_{\text{initial}}| \leq 0.01 \right\}
\]
where \(c(x,y)\) denotes the predicted species concentration at spatial location \((x,y)\), \(c_{\text{ideal}}\) represents the initial (noiseless) concentration, and \(\partial \Omega_{\text{uc}}^{(k)}\) is the set of points located at the immediate boundary of the previously identified region \(\Omega_{\text{uc}}^{(k)}\). By repeatedly applying this criterion, the algorithm progressively moves upward through the spatial domain until reaching the corroded interface region. Conveniently, points within the corroded region exhibit sufficiently distinct concentration values, enabling a clear and robust delimitation between corroded and uncorroded areas. Finally, all points identified within the uncorroded region have their species concentration values replaced with the corresponding ideal, noise-free concentration values, ensuring consistency and adherence to physical constraints. A sketch of the correction algorithm is shown in Figure~\ref{fig:masking}.

\subsection{Initial Condition Generation}
\label{initial_cond}

In this section, we outline how the initial conditions are generated using diffusion models.

\subsubsection{Random Initial Conditions}

In the liquid metal dealloying simulations considered in this study, the simulations are initialized with a perfectly flat liquid-metal interface. However, in order to produce significant diversity in the computational data for training purposes, we inject low-amplitude noise into the initial species concentration fields, which, due to the highly chaotic nature of the underlying physics, evolve to produce widely varying microstructures over time across multiple runs. On the other hand, the U-Net surrogate is designed to learn the dynamics of deformed and continuously evolving interfaces but lacks the capability to generate the diversity of morphologies when starting from a flat interface, even with noise injection. Consequently, it is necessary to initialize surrogate simulations with an already deformed interface, even if the deformation is slight. While such deformed initial conditions can be generated by running the direct numerical solver for a short duration, as done in \cite{bonneville_2025}, this approach is computationally expensive and time-consuming. Given the chaotic nature of the physics, these deformed initial conditions can be viewed as random samples drawn from an unknown probability distribution. To efficiently generate these random initial conditions, we propose training a generative model for images, which can produce diverse deformed interfaces that serve as suitable starting points for surrogate simulations. 

\subsubsection{Diffusion Models}
When properly trained, generative models for computer vision applications are able to generate random, yet visually realistic, images. The key idea of most generative approaches is to learn a mapping between a known, easy-to-sample distribution (e.g. Gaussian) and an approximation of the probability distribution of the training data (or a simpler latent representation). Earlier popular generative models include Restricted Boltzmann Machines~\cite{10.5555/104279} and Deep Belief Networks~\cite{Hinton:2006}. Over the past decade, with the rise of deep learning, Variational Auto-Encoder (VAE)~\cite{Kingma:2022}, Normalizing Flows~\cite{Rezende:2016}, Generative-Adversarial-Networks (GAN) \cite{NIPS2014_5ca3e9b1} and Denoising Diffusion Probabilistic Model (DDPM) \cite{Ho:2020} methods have been proposed. In fact, DDPM, often simply referred to as a \textit{diffusion model} has recently found tremendous utility in multi-modal AI applications \cite{Ramesh:2021,Rombach:2021}. In this paper, we rely on diffusion models, as they can generate high quality field images and are reasonably easy to train and implement. In contrast, GANs are notoriously difficult to train~\cite{Salimans:2016,Mescheder:2018,Che:2017} and can suffer from mode collapse (i.e. degradation of the diversity in the generated data)~\cite{Che:2017,Xu:2021}, and VAEs tend to generate overly blurry images~\cite{Rosca:2017}.

The key idea of diffusion models is to learn a denoising mapping to progressively convert a latent random variable into an image. Such a mapping can be learned by considering two stochastic processes, the \textit{forward} diffusion process $q$, and the \textit{backward} diffusion process, $p$. We consider a set of latent tensors $\{\mathbf{x}_t\}_{t=0}^T$, where $\mathbf{x}_0$ is an original field image from the training dataset. During the forward process, Gaussian noise is added progressively to $\mathbf{x}_0$, over $T$ steps, until it effectively becomes a tensor $\mathbf{x}_T$ of pure noise (and where each of the $\mathbf{x}_t$ represents an increasingly noisy step, for $t\in[\![1,T-1]\!]$). In the forward process, the transition probability from $\mathbf{x}_{t-1}$ to $\mathbf{x}_t$ is formally written as:
\begin{equation}
  q(\mathbf{x}_t|\mathbf{x}_{t-1})=\mathcal{N}(\mathbf{x}_t|\sqrt{1-\beta_t}\mathbf{x}_{t-1},\beta_tI),
\end{equation}
where $\beta_t$ is an arbitrary variance schedule, usually chosen to increase as a function of $t$. In the reverse process, the transition probability from $\mathbf{x}_{t}$ to $\mathbf{x}_{t-1}$ is assumed to follow a Gaussian distribution:
\begin{equation}
    p(\mathbf{x}_{t-1}|\mathbf{x}_t)=\mathcal{N}(\mathbf{x}_{t-1}|\mu_\theta(\mathbf{x}_t,t),\Sigma_\theta(\mathbf{x}_t,t))
\end{equation}
The mean $\mu_\theta$ and variance $\Sigma_\theta$ are unknown, but can be approximated with neural networks parameterized on $\theta$. In practice, $\mu_\theta$ is often chosen as a U-Net \cite{Ronneberger:2015}, and it has been shown that a fixed (pre-defined) variance $\Sigma_\theta$ can be sufficient to achieve excellent results~\cite{Ho:2020}. The parameters $\theta$ can be learned by maximizing the log-marginal-likelihood of the reverse process over $\mathbf{x}_0$:

\begin{equation}
\log p_\theta(\mathbf{x}_0) = \log \int p_\theta(\mathbf{x}_0,\mathbf{x}_1,\dots,\mathbf{x}_T)d\mathbf{x}_1\dots d\mathbf{x}_T
\end{equation}
where $p_\theta(\mathbf{x}_0,\mathbf{x}_1,\dots,\mathbf{x}_T)$ is the joint distribution of the data and latent variables. Unfortunately, directly computing $\log p_\theta(\mathbf{x}_0)$ is generally intractable. Instead, variational inference is used to derive a tractable lower bound on the log-likelihood, known as the Evidence Lower Bound (ELBO) \cite{Ho:2020}. Through Jensen's inequality, it can be shown that:
\begin{equation}
-\log p_\theta(\mathbf{x}_0) \leq \mathcal{L}_{\text{ELBO}}
\end{equation}
with
\begin{equation}
\label{lelbo}
\mathcal{L}_{\text{ELBO}} = -\mathbb{E}_q \left[ \sum_{t=1}^T D_{\text{KL}} \left( q(\mathbf{x}_{t-1}|\mathbf{x}_t, \mathbf{x}_0) \parallel p_\theta(\mathbf{x}_{t-1}|\mathbf{x}_t) \right) - \log p_\theta(\mathbf{x}_0|\mathbf{x}_1) \right]
\end{equation}
where $D_\text{KL}(q,p)$ is the Kullback–Leibler divergence between $q$ and $p$ \cite{Ho:2020}. During training, in practice, rather than fully computing equation \ref{lelbo}, only a handful of steps $t$ and corresponding noisy images $\mathbf{x}_t$ are randomly selected at each epoch.

As mentioned earlier, diffusion models typically employ U-Nets to approximate the mean of the transition probability, $\mu_\theta$. Conveniently, we can choose to directly re-use a similar U-Net architecture as the one developed for the surrogate model. The only difference is that for the diffusion model, we employ 128 input channels in the first layer, instead of 32 (and double this number at each downsampling layer). Note that the two U-Nets, while having almost the same architecture, are trained independently for different purposes, and thus have a different set of weights. Reusing the same architecture is advantageous as it ensures that samples generated by the diffusion model inherently satisfy periodic boundary conditions, a critical requirement for consistency in the spatial domain. Moreover, since the diffusion models' U-Nets typically require conditioning on the timestep $t$, the surrogate's U-Net built-in conditioning mechanism can be directly utilized. Additionally, the surrogate U-Net's capability to handle conditioning on arbitrary parameters enables the training of a conditional diffusion model that can generate snapshot fields for arbitrary ideal \(c_\text{A}\) inputs, thereby enhancing its flexibility in the current study.


\section{Model Training and Testing}
\label{trainingtesting}

\subsection{Training}

The U-Net surrogate model is trained using a dataset that includes five distinct initial concentrations for species A ($c_A=0.20$, $0.25$, $0.30$, $0.35$, and $0.40$). The training data is generated over a 
$512\times512$ grid and depicts the behavior of species C as it sinks within the A-B alloy until it approaches the lower boundary. It should be noted that for lower values of $c_A$, dealloying occurs at a fast rate such that species C ultimately reaches the lower boundary quicker than cases with higher $c_A$ values. This leads to shorter training simulations and consequently fewer input-output training pairs. To maintain a balanced dataset that contains approximately an equal amount of snapshots for each concentration, additional simulations for lower $c_A$ values are incorporated into the training dataset (a breakdown of the number of snapshots, start and end times, and the number of simulations in the training set is provided in Table~\ref{trainingset}). The model is designed to skip between 50,000 and 100,000 time steps at a time. The surrogate model is implemented using PyTorch \cite{paszke2019pytorch}, and is trained with a standard $\mathcal{L}_2$ relative error loss function (this is a common choice in the neural operator literature \cite{li2021fourier, lu2021deeponet} and has been used in our related earlier work \cite{bonneville_2025}). The surrogate is trained with Adam \cite{kingma2017adam}, with a learning rate of $10^{-4}$ over 80 epochs. The model is trained on a single A100 GPU, and takes about 59 hours to complete training.

The diffusion model is trained on a dataset comprising 10 field snapshots for each of the five $c_A$ values mentioned previously. The snapshots are extracted at $t = 0.5\mu$s for $c_A = 0.2$ and $0.25$, and at $t = 1\mu$s for $c_A = 0.3$, $0.35$, and $0.4$. As a result, the training set consists of only 50 original snapshot fields, but we augment it by leveraging the periodic boundary conditions, and generate proxy field images through random horizontal shifts \cite{Shorten2019ASO}. We also further augment the dataset by flipping the field with respect to the vertical axis, which is reasonable given that the physics involved includes diffusion and reaction, but no advection. Thus, the phase-field physics is expected to appear horizontally isotropic (i.e. the flipped fields are still physically accurate). This augmentation process increases the dataset size by a factor of 20, resulting in a total of 1,000 snapshots. Although the training data is generated on a $512 \times 512$ grid, we downsample it to $256 \times 256$ because training diffusion models on high-resolution images is notoriously challenging and computationally expensive~\cite{Rombach:2021, Chen:2024a}.  The diffusion model is also implemented in PyTorch and is a modified version of Lucidrains' original implementation \cite{lucidrain}. Training is conducted using Adam for 1,000 epochs with a learning rate of $10^{-5}$, requiring about 12 hours to complete on a single GPU.

\begin{table}[ht]
\centering
\begin{tabular}{cccccccc}
\Xhline{3\arrayrulewidth}\\[-6pt]
$c_A$ & \makecell{Initialization\\time} & \makecell{End time\\(training)} & \makecell{End time\\(extrapolation)} & \makecell{Number of snapshot\\per simulation} & \makecell{Number\\of simulations} & \makecell{Total number\\of snapshots}\\[8pt]\hline\\[-6pt]
0.20 & 0.5 & 3.5 & 9.0 & 60 & 21 & 1260\\
0.25 & 0.5 & 4.5 & 12.0 & 80 & 16 & 1280\\
0.30 & 1.0 & 6.0 & 16.0 & 100 & 13 & 1300\\
0.35 & 1.0 & 9.0 & 24.0 & 160 & 8 & 1280\\
0.40 & 1.0 & 9.0 & 24.0 & 160 & 8 & 1280\\
[2pt]\Xhline{3\arrayrulewidth}
\end{tabular}
\vspace{0.1in}
\caption{A breakdown of the simulations contained in the training dataset. Times are given in $\mu$s. The end time correspond to the moment where species C is close to reaching the lower boundary. Each simulation contains snapshots of the field with a 50,000 time step increment. The number of simulations is balanced so that there is roughly 1300 snapshots per concentration values.}
\label{trainingset}
\end{table}

\subsection{Evaluation Metrics and Quantities of Interests}

\textbf{Quantities of Interest:} One of the primary motivations for running phase-field simulations, whether through direct numerical simulation (DNS) or surrogate modeling, is the extraction of quantities of interest (QoIs) that can describe key features of the system state. These QoIs can be used for practical applications that include decision-making, design optimization and/or validation, and uncertainty quantification. In the context of LMD, we focus on a set of QoIs identified as particularly relevant in previous work \cite{McCue_Karma_Erlebacher_2018,Wada:2023,VAKILI20171852,Tran_2019,Lai:2022,MCCUE201610,bonneville_2025}, and that describe the curvature and perimeter of the liquid-metal interface, the penetration depth of species C, the remaining amounts of metal and species A and B, and the average ligament height. These QoIs can be computed at each time step, following the methodology described in \cite{bonneville_2025}. To evaluate the accuracy of our model, we calculate the QoIs based on surrogate predictions at each available time step (e.g., every 50,000 to 100,000 steps) and compare them to DNS-based values by computing the relative $\mathcal{L}_2$ error of the vectorized QoIs \cite{bonneville_2025}:
\begin{equation}
\label{qoi_rel}
e(\pmb{q},\pmb{\hat{q}})=\frac{|\!|\pmb{q}-\pmb{\hat{q}}|\!|_2}{|\!|\pmb{q}|\!|_2}
\end{equation}
where $\pmb{q}=(q(t_\text{start}),\dots,q(t_\text{end}))$ represents a DNS-based QoI vectorized over each time step (and $\pmb{\hat{q}}$ is the equivalent vector computed from the surrogate prediction).

\textbf{Auto-Correlation Errors:} For extremely chaotic systems like LMD, error metrics often used with neural operators or machine learning-based surrogates, such as the relative error between the DNS-based and predicted fields \cite{li2021fourier, lu2021deeponet, kovachki2021neural}, are less informative due to the inherent unpredictability of the dynamics. Instead, comparing invariant statistical properties, such as spatial auto-correlations, provides a more meaningful assessment. In phase-field simulations, spatial auto-correlation quantifies the probability that two points (picked randomly across space) belong to the same phase or species, making it a relevant metric for describing microstructure statistics \cite{li2022learning,HERMAN2020589}. The spatial auto-correlation relative error has been successfully applied in chaotic phase-field models to evaluate surrogate predictions \cite{Oommen:2024, bonneville_2025}. Additionally, for baseline comparison, the average auto-correlation relative discrepancy between members of an ensemble of DNS-based simulations is computed, enabling a meaningful evaluation of surrogate performance against DNS-based simulations with different initial conditions \cite{bonneville_2025}.
Formally, the spatial auto-correlation relative error is:
\begin{equation}
\label{releleq}
    e_\text{AC}(\mathbf{u}, \mathbf{\hat{u}},t)=\frac{|\!|S_{\mathbf{u}\mathbf{u}}(\pmb{r},t)-S_{\mathbf{\hat{u}}\mathbf{\hat{u}}}(\pmb{r},t)|\!|_2}{|\!|S_{\mathbf{u}\mathbf{u}}(\pmb{r},t)|\!|_2}
\end{equation}
where $S_{\mathbf{u}\mathbf{u}}(\pmb{r},t)$ is the 2D spatial auto-correlation of the field $\mathbf{u}$ (representing either the phase or one of the species fields) at time $t$, and $\pmb{r}$ is a vector between any two random spatial locations. 

\subsection{Testing}

Surrogate simulations are conducted using an auto-regressive roll-out approach, where the surrogate model iteratively predicts the field at successive time steps. Starting from an initial condition, the surrogate output at a later time is fed back into the model as input, allowing predictions to extend to an arbitrary time horizon \cite{bonneville_2025, Oommen:2024}. Initial conditions are selected based on the field at $t = 0.5\mu$s (for $c_A = 0.20$ and $c_A = 0.25$) or $t = 1\mu$s (for $c_A = 0.3$, $c_A = 0.35$, and $c_A = 0.4$), and are generated using either the numerical solver (by running a short simulation from $t = 0\mu$s to $t = 0.5\mu$s or $1\mu$s) or the diffusion model. As outlined in section \ref{unet-arch}, the U-Net surrogate is purely convolution-based, and thus independent of input dimensions so we can run surrogate simulations over an extended domain twice as tall as the training domain (i.e. a 1024$\times$512 grid). This extended domain allows species C to penetrate deeper into the A-B alloy, and we extrapolate the surrogate predictions by rolling out simulations for a time horizon approximately three times longer than the training data end time for each concentration. This extrapolation end time is chosen such that species C can reach close to the lower boundary of the extended domain, and a specific breakdown of these end times for each concentration is provided in table \ref{trainingset}. The extrapolation domain size is chosen to balance computational feasibility. Indeed, generating DNS-based test simulations for comparison purposes over a 1024$\times$512 grid can take up to several months on 64 to 128 cores, so an even larger domain would require prohibitively long runtimes. To evaluate surrogate accuracy, we compute the QoIs and spatial auto-correlation relative errors across both the training regime and the extrapolation regime. Note that here, the term \textit{training regime} only refers to time spans that match the ones included in the training dataset, but the test surrogate simulations are compared against DNS-based simulations that are completely independent of the training data (including in the \textit{training regime}).

We evaluate the surrogate model using initial conditions generated either by the numerical solver or by the diffusion model. In the latter case, the initial conditions are sampled from a purely data-driven generative model, with little guarantee that the generated samples rigorously respect the physical governing equations. Properly assessing the quality of samples produced by generative models remains a challenging and open question in the machine learning community~\cite{Theis:2016, Salimans:2016, Borji:2018}. This is particularly difficult here, because we are generating abstract physical fields rather than real-life images, so a visual inspection is a very poor validation metric. Ultimately, the best validation test lies in the downstream task: When initialized with generative samples, can the surrogate model capture the phase-field dynamics (i.e. QoIs, auto-correlation errors, etc.) as accurately as when initialized with the numerical solver? For an initial sanity check however, we estimate the marginal densities of the QoIs over the generated samples and compare them to the estimated densities of the QoIs derived from the DNS-based (solver-generated) test data. This comparison involves both visual inspection of the density distributions and a Kolmogorov-Smirnov (KS) test \cite{kolmogorov_1951} to quantitatively assess, for each available concentration and QoI, how well the distributions align. 

\begin{figure}[!ht]
\centering
    \includegraphics[width=1.\textwidth]{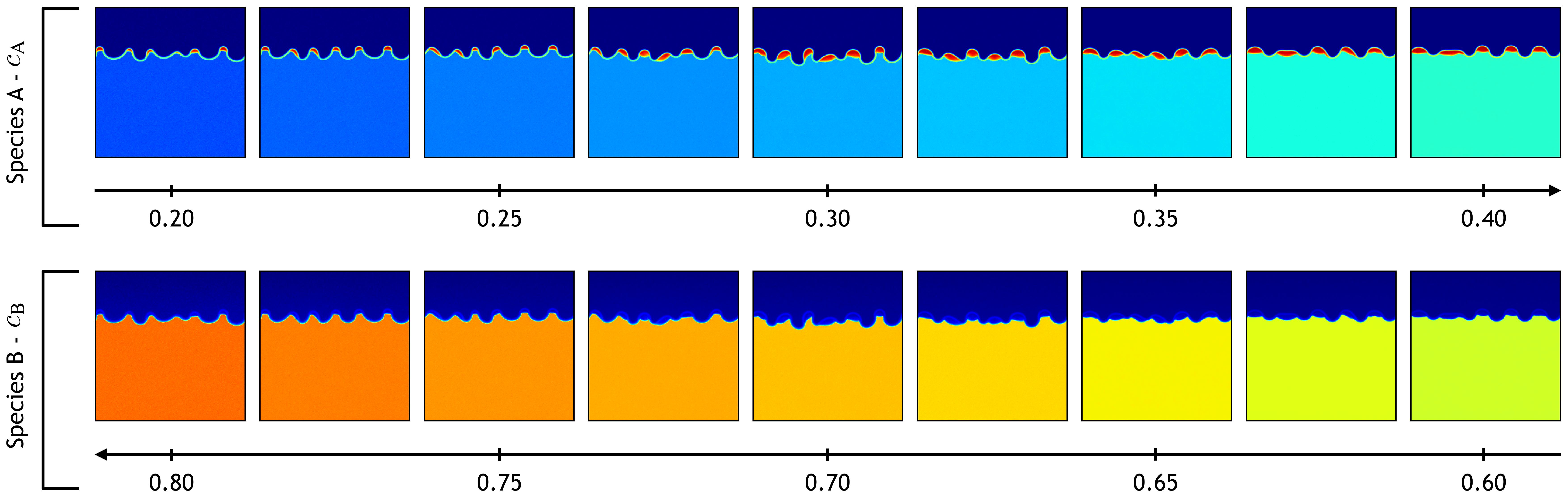}
    \caption{Example of initial conditions sampled from the diffusion model. The diffusion model is conditioned on $c_A$, and can thus be used to generate fields for arbitrary values of $c_A$ not contained in the training dataset (e.g. $c_A=0.225$). In this figure, samples for $c_A$ ranging from 0.2 to 0.4, with a 0.025 increment are shown (and corresponding $c_B=1-c_A$).}
    \label{fig:samples}
\end{figure}

\begin{figure}[!ht]
\centering
    \includegraphics[width=1.\textwidth]{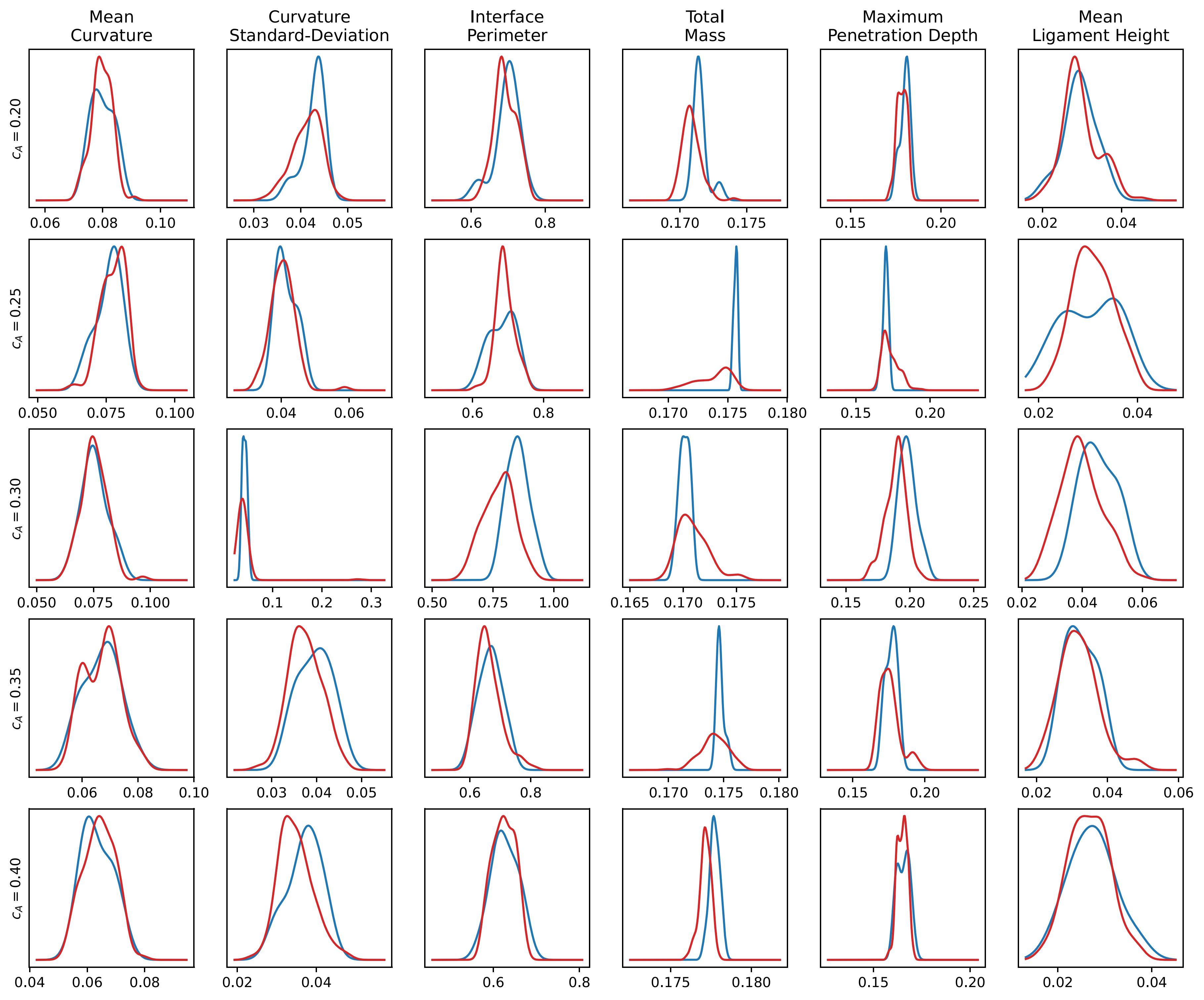}
    \caption{QoI marginal densities for different values of $c_A$. Results in red (\raisebox{1.5pt}{\protect\tikz{\protect\draw[default line, very thick, line cap=round, Color2] (0, 0.) -- (0.33, 0.);}}) correspond to diffusion model-generated samples, and blue lines (\raisebox{1.5pt}{\protect\tikz{\protect\draw[default line, very thick, line cap=round, Color1] (0, 0.) -- (0.33, 0.);}}) are for DNS-based (solver-generated) samples. Empirically, the distributions match reasonably well, except for the total mass QoI (as seen in table \ref{tab:pval}). However, unlike other QoIs, the total mass has an extremely reduced range (typically between 0.165 and 0.18), so the diffusion model still generates samples that are on average close to the true distribution. The distributions are estimated using kernel density estimation (KDE) \cite{chen2017tutorialkerneldensityestimation}.}
    \label{fig:gen_qoi_distrib}
\end{figure}

The KS test is a statistical method used to determine whether samples from two distributions could have been drawn from the same underlying distribution. Unlike other similar tests, the KS test does not assume any specific form for the underlying distribution (e.g., it does not require the distributions to be Gaussian). Figure \ref{fig:samples} shows field samples generated with the diffusion model, for a range of values of $c_A$. Figure \ref{fig:gen_qoi_distrib} shows the QoI distributions based on both solver and diffusion model-generated fields, and table \ref{tab:pval} shows the corresponding KS test $p$-values for each concentration and QoI label. The $p$-value is a metric used to quantify the likelihood that two sets of random samples are drawn from the same probability distribution \cite{Fisher_1925, fisher1958statistical}. In general, we observe a reasonable visual match between the distributions, and $p$-values from the KS test often exceed 0.05, indicating close agreement between the distributions. For the total mass of metal, however, the synthetic samples slightly underestimate the amount of metal alloy, with the mean shifted downward (though not significantly) and the distribution spread slightly overestimated. Additionally, the diffusion model appears to struggle more with $c_A = 0.3$, as other QoIs, such as the maximum penetration depth and the mean ligament height, also yield $p$-values below 0.05, suggesting slightly less accurate distribution matches. 

Since the diffusion model was trained on downsampled 256$\times$256 field snapshots, the generated samples retain these dimensions. To adapt them for the surrogate simulations, we upsample the snapshots back to 512$\times$512 using linear interpolation. For predictions over the taller 1024$\times$512 domain, we concatenate the upsampled samples with an additional 512$\times$512 tensor representing pure metal, where $\phi = 1$  and $c_A $ and $c_B $ are uniformly constant throughout the domain.

\begin{table}[ht]
\centering
\begin{tabular}{ccccccc}
\Xhline{3\arrayrulewidth}\\[-6pt]
$c_A$ & \makecell{Mean\\Curvature} & \makecell{Curvature\\ Standard Deviation} & \makecell{Interface\\ Perimeter} & Total Mass & \makecell{Maximum\\Penetration Depth} & \makecell{Mean Ligament\\ Height}\\[8pt]\hline\\[-6pt]
0.20 & 0.6090 & 0.1157 & 0.1777 & \textbf{0.0001} & 0.2315 & 0.9274\\
0.25 & 0.3326 & 0.6090 & 0.2624 & \textbf{0.0000} & 0.0520 & 0.4605\\
0.30 & 0.9865 & 0.0851 & \textbf{0.0018} & \textbf{0.0171} & \textbf{0.0171} & \textbf{0.0365}\\
0.35 & 0.9999 & 0.2315 & 0.4605 & \textbf{0.0059} & 0.3326 & 0.9274\\
0.40 & 0.6604 & 0.1157 & 0.9735 & \textbf{0.0000} & 0.9274 & 0.9274\\
[2pt]\Xhline{3\arrayrulewidth}
\end{tabular}
\vspace{0.1in}
\caption{$p$-values for the Kolmogorov-Smirnov test between the QoI distributions of the solver-generated samples and the diffusion model-generated samples. The diffusion model-generated samples appear to statistically match well true samples, except for the total mass (for this QoI, most $p$-values are below 0.05, as highlighted in bold).}
\label{tab:pval}
\end{table}

\FloatBarrier

\section{Results}
\label{results}

We now discuss the performance of our surrogate model, with and without generative initial conditions. Note that unless explicitly mentioned otherwise, the surrogate model employs all the components outlined in section \ref{surrogate}, and skips $50,000$ time steps at each forward pass. In section \ref{sec:comp}, we specifically discuss the effect of tweaking certain model components.

\subsection{Surrogate Performance with Solver-Generated Initial Conditions}

\begin{figure}[!htb]
\centering
    \includegraphics[width=1.\textwidth]{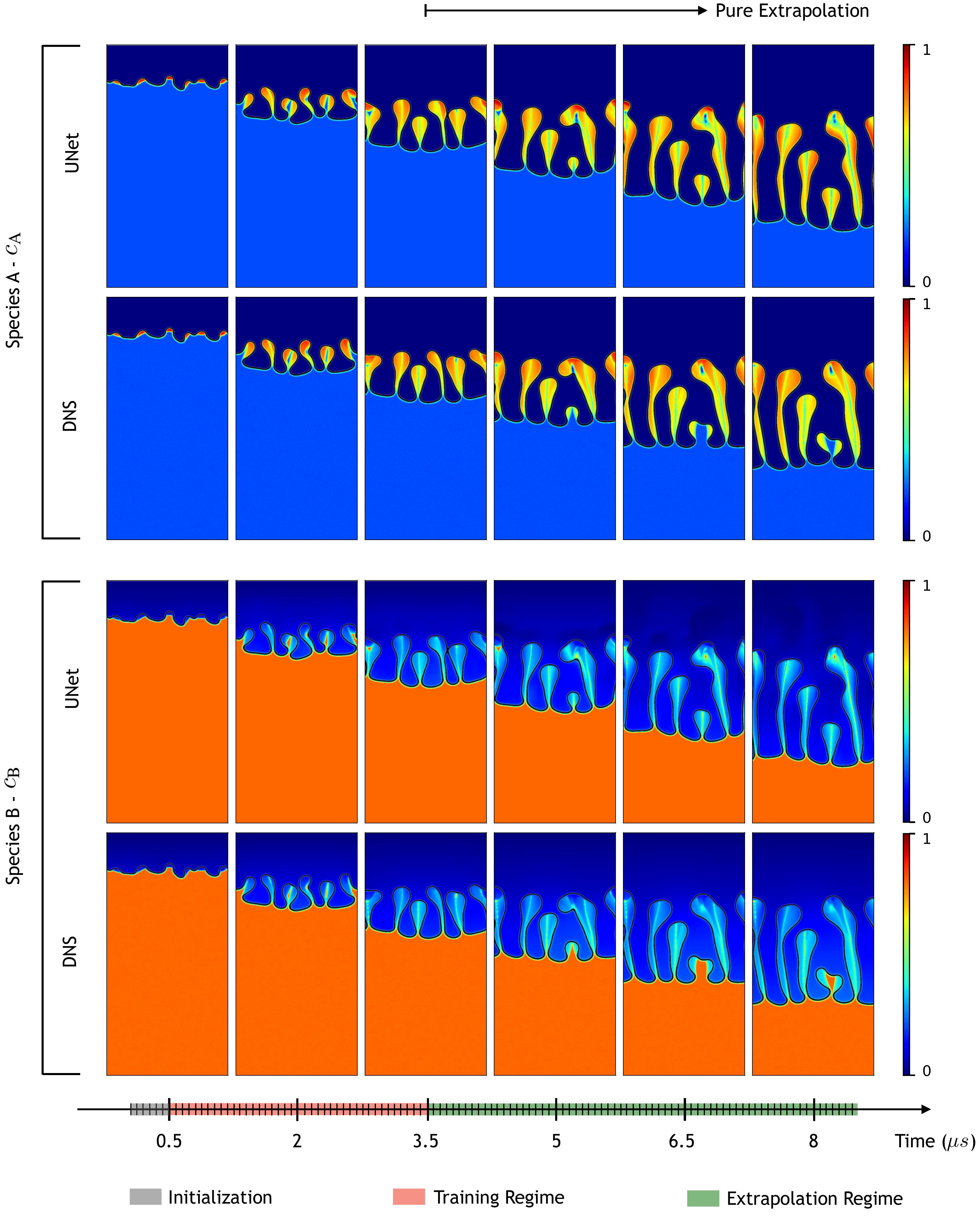}
    \caption{Concentration fields for species A and B. The ideal (pre-dealloying) concentration for A is $\pmb{c_A=0.2}$. For each species, the top row represents the U-Net surrogate prediction, and the lower row is the DNS-based solution, obtained from the same initial condition through direct numerical simulation (DNS). The black contours (\raisebox{1.5pt}{\protect\tikz{\protect\draw[default line, very thick, line cap=round, Color3] (0, 0.) -- (0.33, 0.);}}) represent the limit of the liquid-metal interface, as given by the binary phase field $\phi$.}
    \label{fig:field_020}
\end{figure}

\begin{figure}[!htb]
\centering
    \includegraphics[width=1.\textwidth]{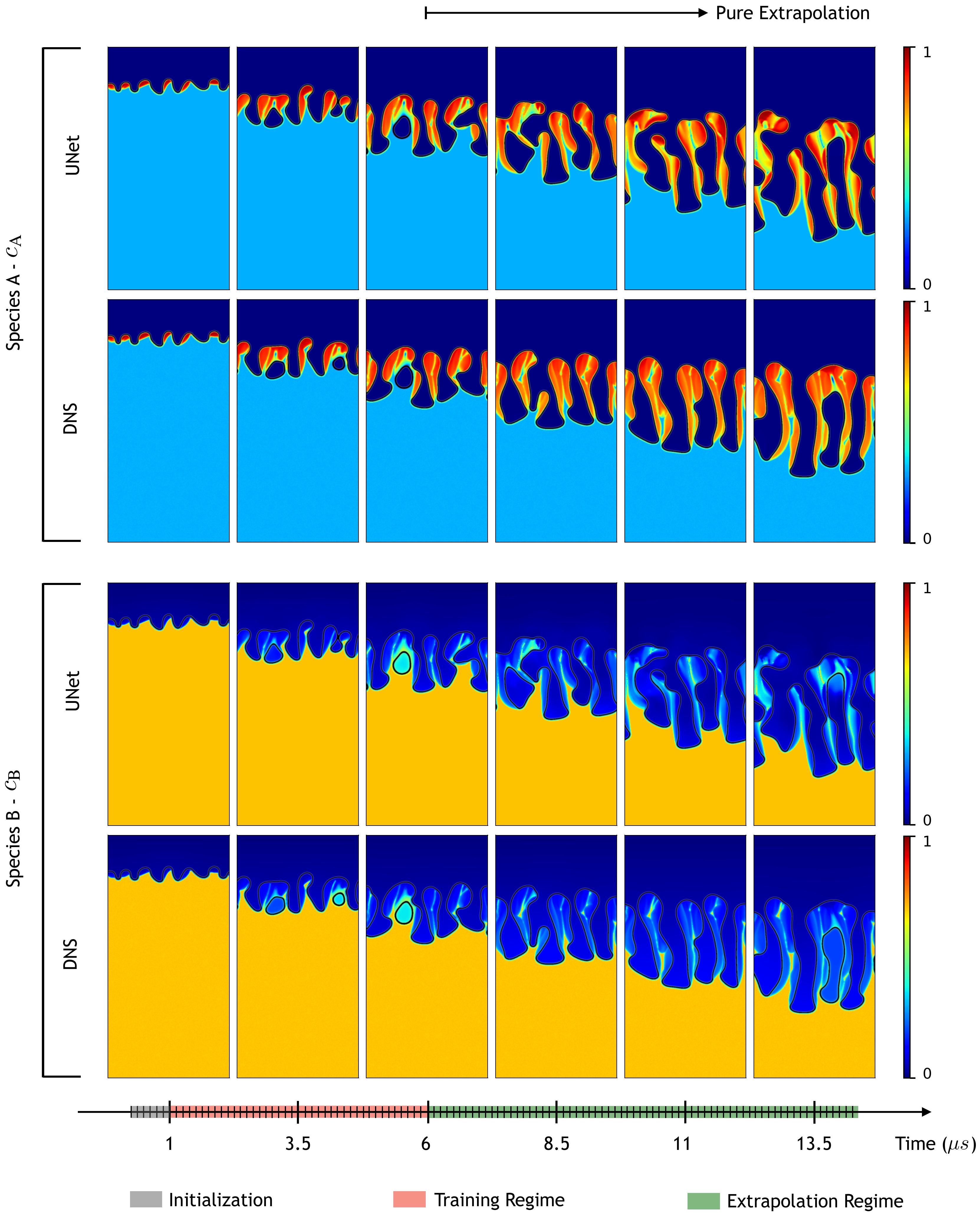}
    \caption{Concentration fields for species A and B. The ideal (pre-dealloying) concentration for A is $\pmb{c_A=0.3}$.}
    \label{fig:field_030}
\end{figure}

\begin{figure}[!htb]
\centering
    \includegraphics[width=1.\textwidth]{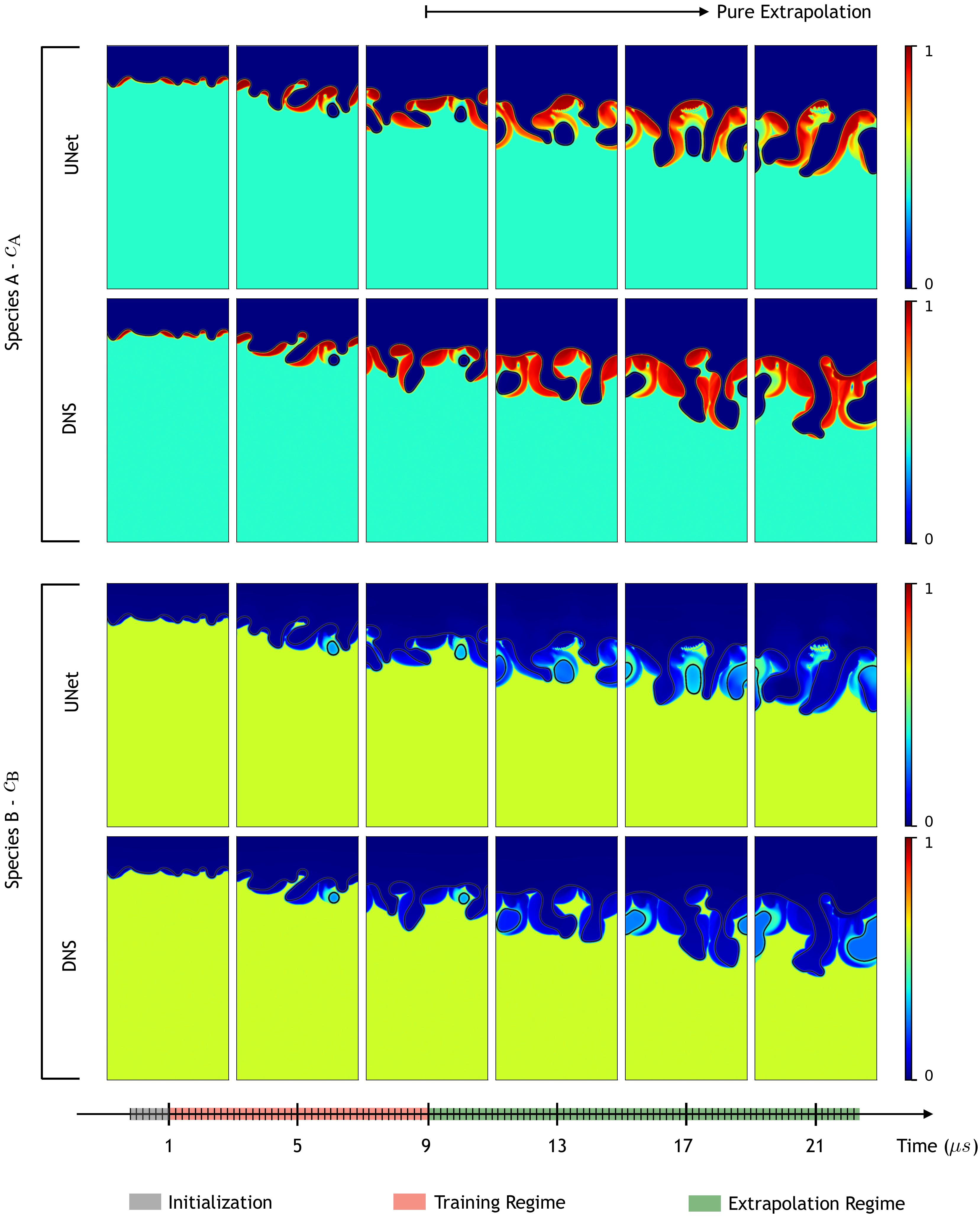}
    \caption{Concentration fields for species A and B. The ideal (pre-dealloying) concentration for A is $\pmb{c_A=0.4}$.}
    \label{fig:field_040}
\end{figure}

Figures \ref{fig:field_020}, \ref{fig:field_030} and \ref{fig:field_040} show examples of test simulations conducted using the U-Net surrogate model applied to \textbf{solver-generated} initial conditions for ideal species concentrations $c_A = 0.2$, $c_A = 0.3$, and $c_A = 0.4$, respectively. Each figure presents both the U-Net predictions and the corresponding true solutions, showcasing the evolution of species concentrations $c_A$ and $c_B$. The interface of the binary phase field $\phi$ is superposed as a black contour to highlight morphological features. Given the highly chaotic nature of the phase field dynamics, the microstructure morphologies eventually exhibit slight mismatches with the DNS-based as the simulations progress, which is both expected and inevitable. Yet, the overall physical features of the time evolving structure of the system appear to be well captured. As species $C$ sinks into the $A$-$B$ alloy, the interface shifts downward at an appropriate pace. The curvature of the interface aligns well with the true fields, and the dendrite shapes and sizes also appear to evolve accurately. The surrogate exhibits the growing de-alloying patterns at the interface, characterized by spikes in $A$ concentration near the interface and the complete erosion of $B$ species. Notably, the surrogate simulation remains stable, showing no signs of un-physical artifacts or numerical instabilities. As the model enters the extrapolation regime, the predictions keep unrolling smoothly, with no apparent drop in the physical fidelity of the predicted fields.

\begin{figure}[!ht]
\centering
    \includegraphics[width=1.\textwidth]{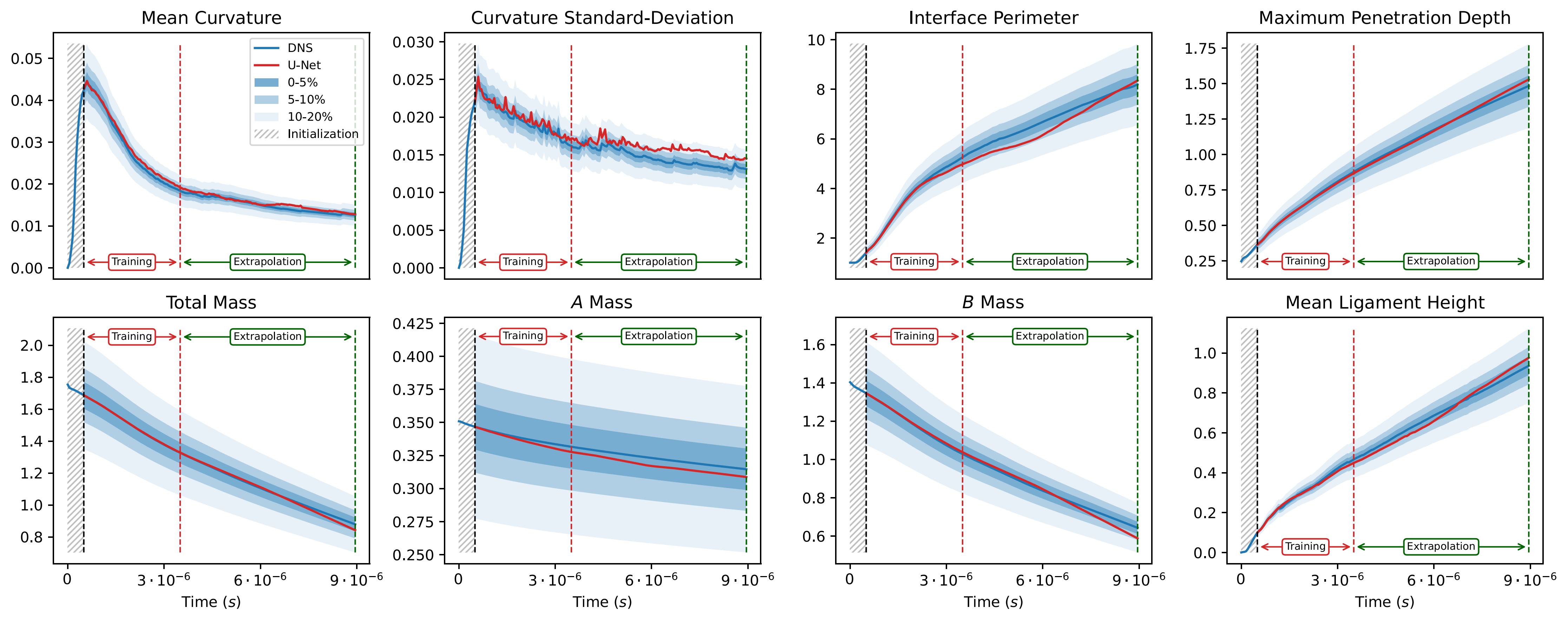}
    \caption{Mean QoI predictions for both the U-Net surrogate, in red (\raisebox{1.5pt}{\protect\tikz{\protect\draw[default line, very thick, line cap=round, Color2] (0, 0.) -- (0.33, 0.);}}) and the numerical solver, in blue (\raisebox{1.5pt}{\protect\tikz{\protect\draw[default line, very thick, line cap=round, Color1] (0, 0.) -- (0.33, 0.);}}). The ideal species is $\pmb{c_A=0.2}$. The relative error ranges are shown in shaded blue. The extents of the training and extrapolation regimes are shown in red and green, respectively.}
    \label{fig:qoi_020}
\end{figure}

\begin{figure}[!ht]
\centering
    \includegraphics[width=1.\textwidth]{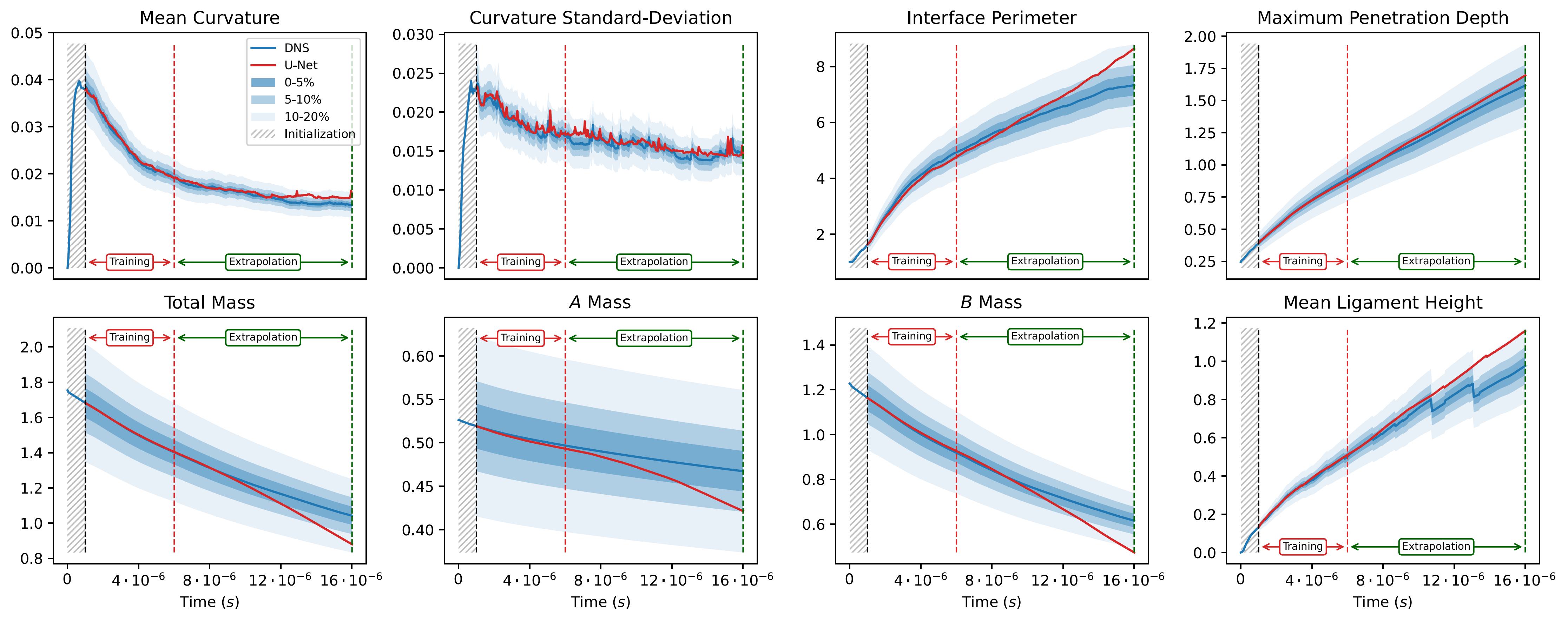}
    \caption{Mean QoI predictions for both the U-Net surrogate, in red (\raisebox{1.5pt}{\protect\tikz{\protect\draw[default line, very thick, line cap=round, Color2] (0, 0.) -- (0.33, 0.);}}) and the numerical solver, in blue (\raisebox{1.5pt}{\protect\tikz{\protect\draw[default line, very thick, line cap=round, Color1] (0, 0.) -- (0.33, 0.);}}). The ideal species is $\pmb{c_A=0.3}$. In the extrapolation regime, the surrogate model increasingly underestimates the amount of A-B alloy remaining (total mass), and both A and B masses. Note that the step-like discontinuities occurring with the solver for the mean ligament height is due to the way the QoI is computed: The mean ligament height measures the average height between the deepest penetration point and the highest tip of metal dendrites \textbf{still attached to the bulk} of the alloy. Since it is possible for large chunks of metal to suddenly detach (\textit{pinch-off effect}), the height of the dendrite tips can change suddenly.}
    \label{fig:qoi_030}
\end{figure}

\begin{figure}[!ht]
\centering
    \includegraphics[width=1.\textwidth]{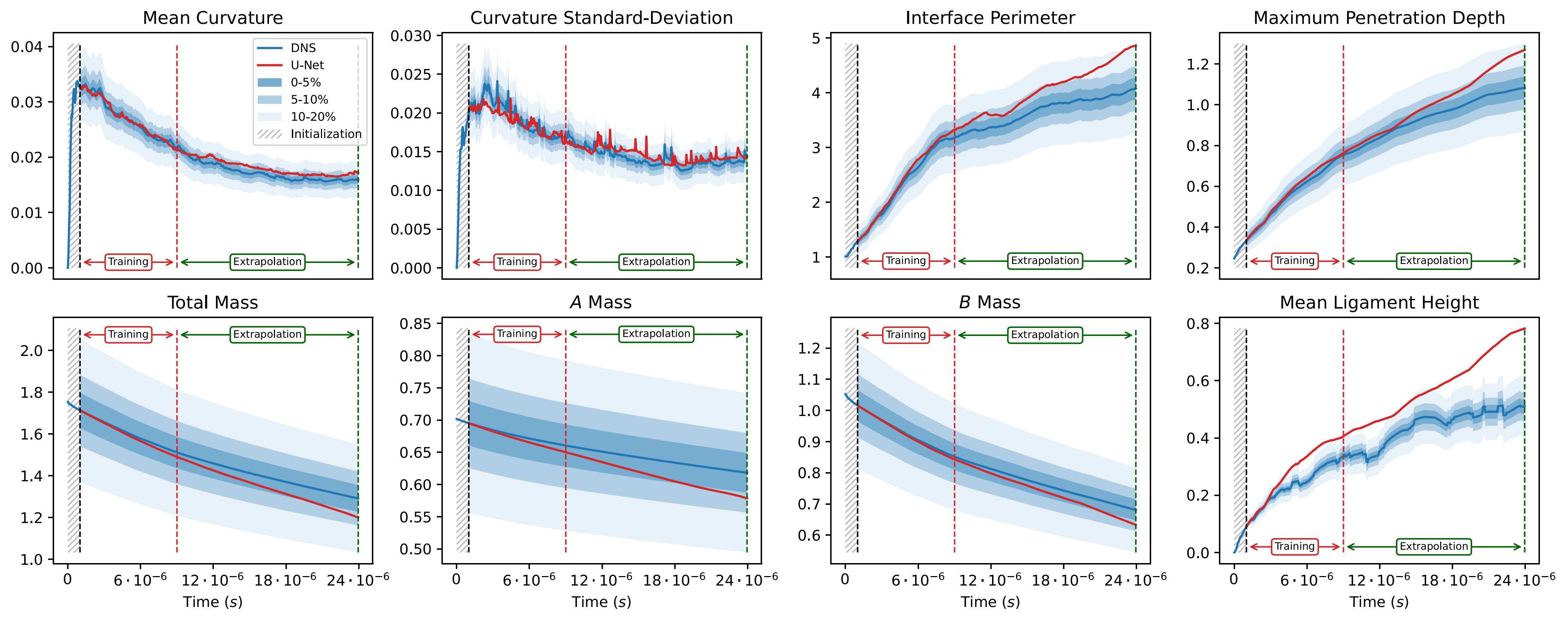}
    \caption{Mean QoI predictions for both the U-Net surrogate, in red (\raisebox{1.5pt}{\protect\tikz{\protect\draw[default line, very thick, line cap=round, Color2] (0, 0.) -- (0.33, 0.);}}) and the numerical solver, in blue (\raisebox{1.5pt}{\protect\tikz{\protect\draw[default line, very thick, line cap=round, Color1] (0, 0.) -- (0.33, 0.);}}). The ideal species is $\pmb{c_A=0.4}$. Here, the surrogate somewhat struggles to capture the right penetration rate and mean ligament height evolution in the extrapolation. The true de-alloying process clearly slows down, whereas the surrogate's pace is roughly linear (for the penetration depth over time).}
    \label{fig:qoi_040}
\end{figure}

\begin{figure}[!ht]
\centering
    \includegraphics[width=1.\textwidth]{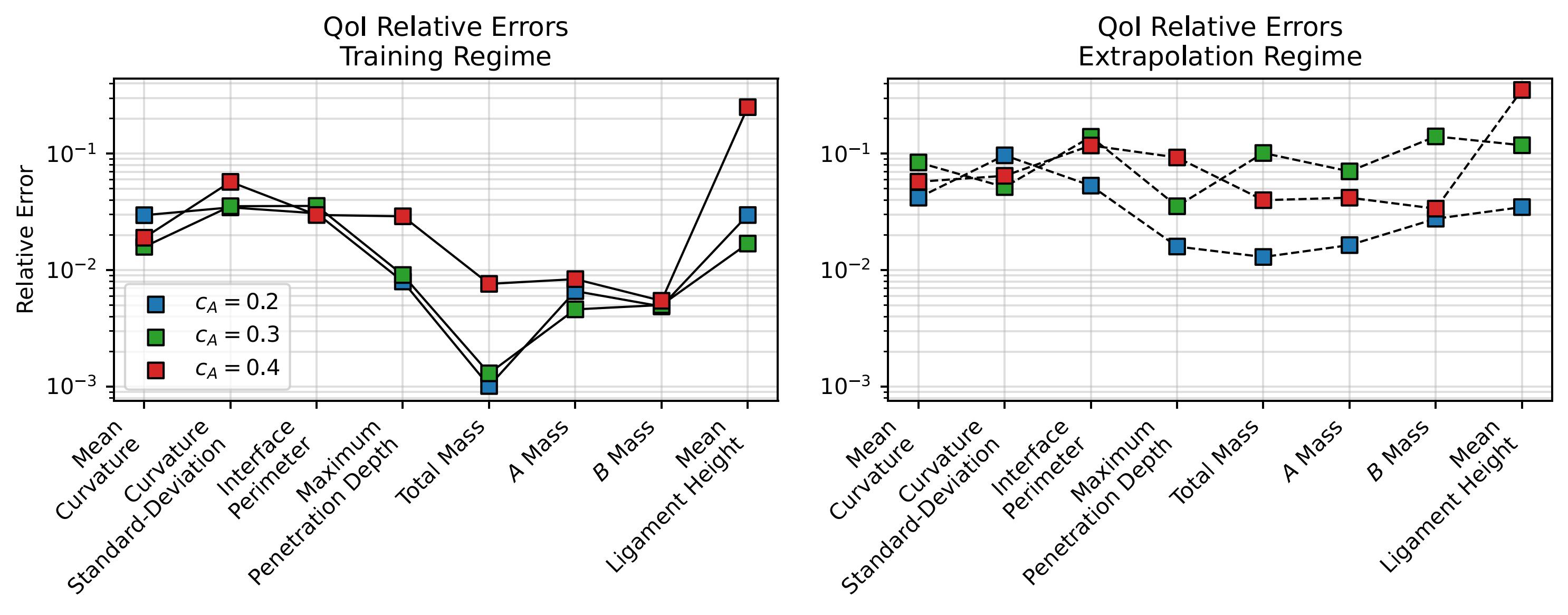}
    \caption{QoI relative errors for both the training and the extrapolation regime, as defined in equation \ref{qoi_rel}.}
    \label{fig:qoi_errors_allca}
\end{figure}

Figures \ref{fig:qoi_020}, \ref{fig:qoi_030}, and \ref{fig:qoi_040} present the time evolution of the eight different QoIs considered in this paper for species concentrations $c_A = 0.2$, $c_A = 0.3$, and $c_A = 0.4$, respectively. Each figure illustrates the evolution of the QoIs over time, based on both the U-Net surrogate predictions and the DNS-based numerical simulations, averaged over 10 independent test simulations. Both the surrogate and DNS-based simulations begin from identical solver-generated initial conditions. For convenience, relative error ranges (compared to the averaged DNS-based) are also provided in the figures. For all concentrations, the surrogate model predictions remain closely aligned with the DNS-based, particularly in the training regime, where errors typically fall within the 5–10\% range. In the extrapolation regime, the surrogate begins to accumulate some error buildup, especially for higher $c_A$ concentrations, although the errors generally remain within reasonable bounds (less than 20\%). These QoI predictions are consistent with the earlier empirical observation that the underlying physics is (reasonably) well respected. Figure \ref{fig:qoi_errors_allca} provides a detailed breakdown of relative errors for each QoI and $c_A$ values, distinguishing between the training and extrapolation regimes. The surrogate model performs notably better for lower $c_A$ values, as for $c_A = 0.2$ errors in the training regime remain below 3–4\%, and even in the extrapolation regime, they stay under 10\%. For higher $c_A$ values though, the accuracy slightly deteriorates, with errors generally below 10\% in the training regime and 15–20\% in extrapolation. Among all the QoIs, the mean ligament height is the most challenging to predict accurately, particularly for $c_A = 0.4$. This may be due to the fact that, as seen in figure \ref{fig:qoi_040}, the DNS-based de-alloying process slows down only after exiting the training regime. The model may not have been exposed to certain physical patterns that emerge later, making them inherently difficult to capture.

\begin{figure}[!ht]
\centering
    \includegraphics[width=1.\textwidth]{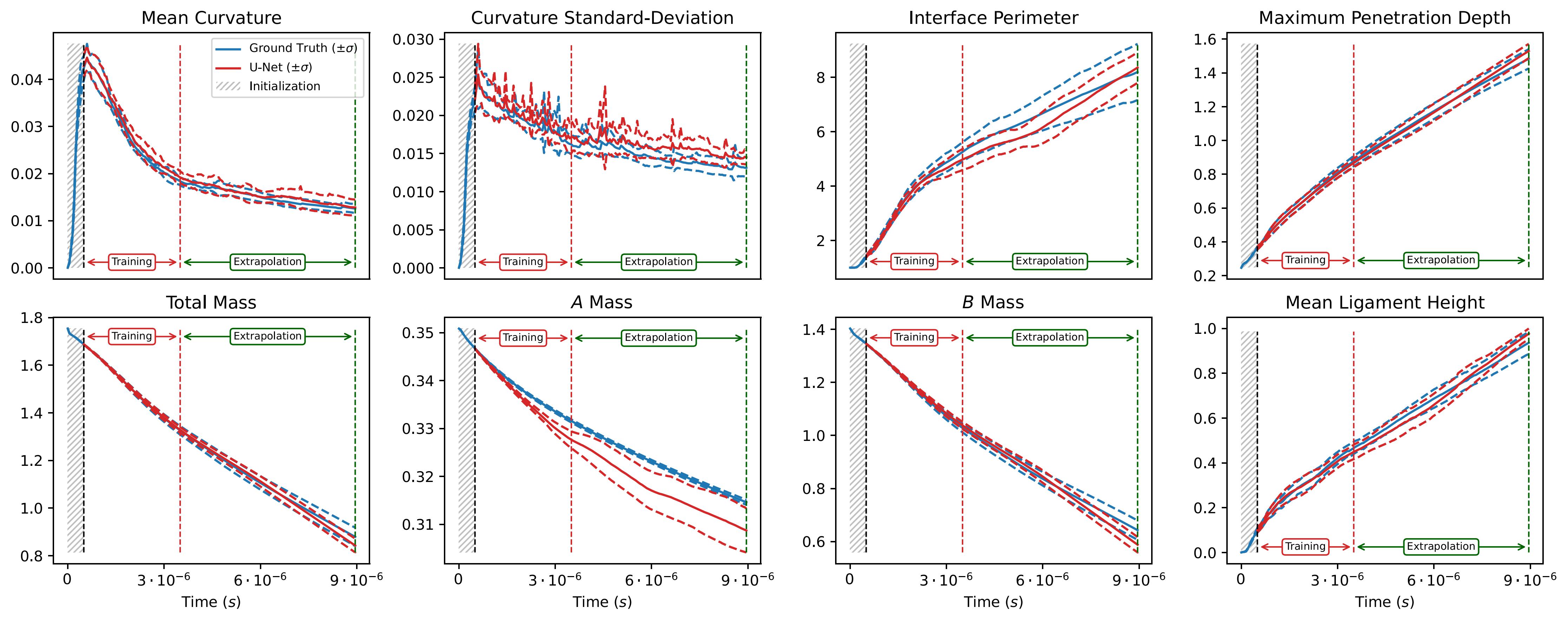}
    \caption{QoI mean and standard-deviation predictions for both the U-Net surrogate, in red (\raisebox{1.5pt}{\protect\tikz{\protect\draw[default line, very thick, line cap=round, Color2] (0, 0.) -- (0.33, 0.);}}) and the numerical solver, in blue (\raisebox{1.5pt}{\protect\tikz{\protect\draw[default line, very thick, line cap=round, Color1] (0, 0.) -- (0.33, 0.);}}). The ideal species is $\pmb{c_A=0.2}$.}
    \label{fig:qoi_std_20}
\end{figure}

Figure \ref{fig:qoi_std_20} presents results similar to those shown in Figure \ref{fig:qoi_020} but now includes the standard deviation of each QoI prediction (rather than just the mean). The standard deviations represent the fluctuations observed in the QoIs across the 10 test simulations, providing insight into the variability of the surrogate and DNS-based predictions. Remarkably, the surrogate is able to capture approximately the correct standard deviation for most QoIs, indicating that it can reproduce the same diversity of simulations as observed in the DNS-based. This suggests that the surrogate model is not only accurate in predicting the mean behavior but also capable of reflecting the inherent variability in the physical system. The only notable exception is the A mass, for which the surrogate exhibits a  broader spread than expected.

\begin{figure}[!ht]
\centering
    \includegraphics[width=1.\textwidth]{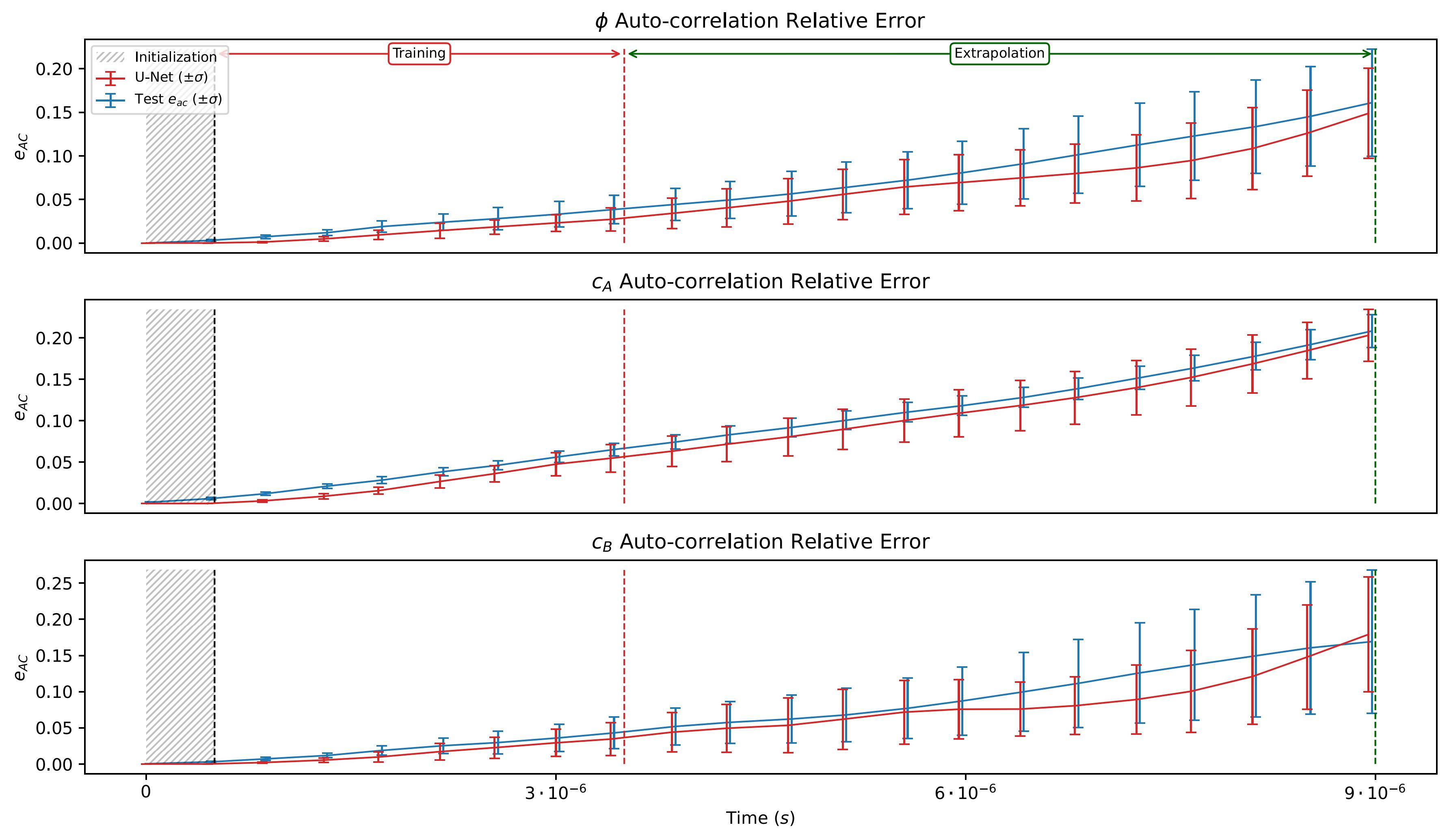}
    \caption{Auto-correlation relative errors for $\pmb{c_A=0.2}$. Red lines (\raisebox{1.5pt}{\protect\tikz{\protect\draw[default line, very thick, line cap=round, Color2] (0, 0.) -- (0.33, 0.);}}) represent the mean auto-correlation relative error achieved with the U-Net surrogate. Blue lines (\raisebox{1.5pt}{\protect\tikz{\protect\draw[default line, very thick, line cap=round, Color1] (0, 0.) -- (0.33, 0.);}}) represent the average auto-correlation relative \textit{discrepancy} between any two (DNS-based) test simulations. For enhanced readability, the standard deviation intervals of the test set discrepancy are slightly shifted with respect to the ones achieved with the U-Net surrogate.}
    \label{fig:ac_020}
\end{figure}

\begin{figure}[!ht]
\centering
    \includegraphics[width=1.\textwidth]{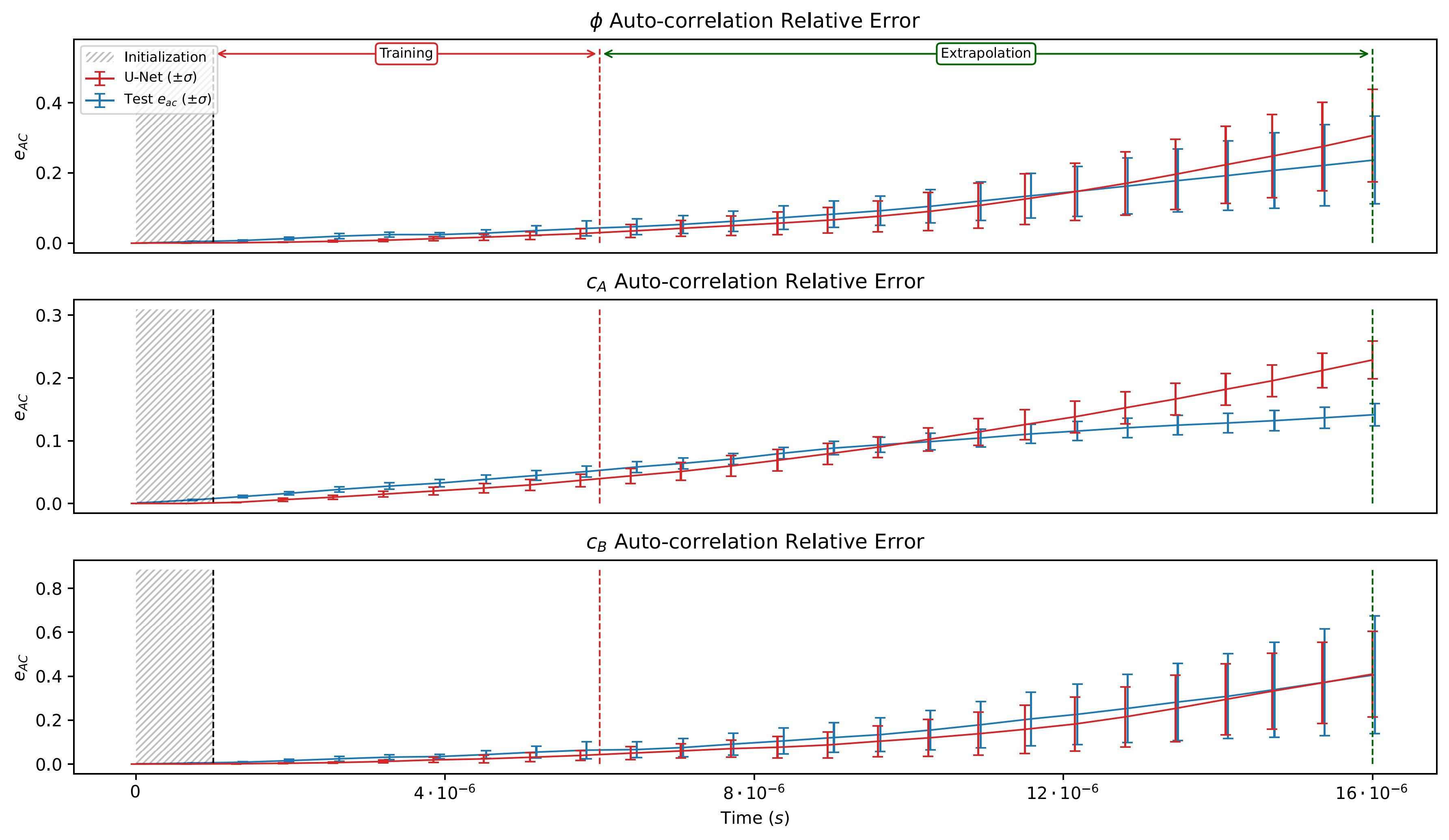}
    \caption{Auto-correlation relative errors for $\pmb{c_A=0.3}$. Red lines (\raisebox{1.5pt}{\protect\tikz{\protect\draw[default line, very thick, line cap=round, Color2] (0, 0.) -- (0.33, 0.);}}) represent the mean auto-correlation relative error achieved with the U-Net surrogate. Blue lines (\raisebox{1.5pt}{\protect\tikz{\protect\draw[default line, very thick, line cap=round, Color1] (0, 0.) -- (0.33, 0.);}}) represent the average auto-correlation relative \textit{discrepancy} between any two (DNS-based) test simulations.}
    \label{fig:ac_030}
\end{figure}

\begin{figure}[!ht]
\centering
    \includegraphics[width=1.\textwidth]{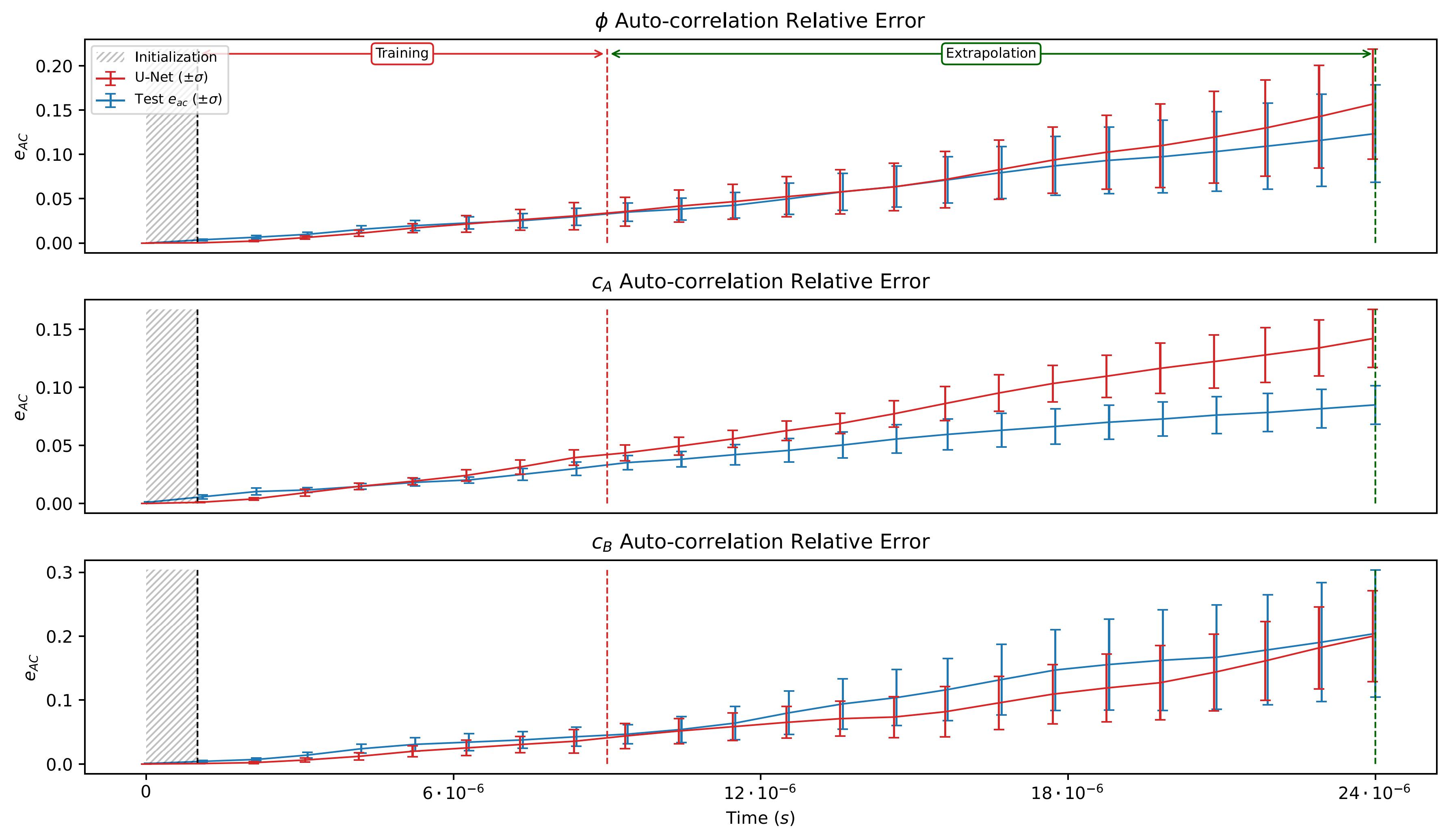}
    \caption{Auto-correlation relative errors for $\pmb{c_A=0.4}$. Red lines (\raisebox{1.5pt}{\protect\tikz{\protect\draw[default line, very thick, line cap=round, Color2] (0, 0.) -- (0.33, 0.);}}) represent the mean auto-correlation relative error achieved with the U-Net surrogate. Blue lines (\raisebox{1.5pt}{\protect\tikz{\protect\draw[default line, very thick, line cap=round, Color1] (0, 0.) -- (0.33, 0.);}}) represent the average auto-correlation relative \textit{discrepancy} between any two (DNS-based) test simulations.}
    \label{fig:ac_040}
\end{figure}

Figures \ref{fig:ac_020}, \ref{fig:ac_030}, and \ref{fig:ac_040} present the relative auto-correlation errors for $c_A = 0.2$, $c_A = 0.3$, and $c_A = 0.4$, respectively. In these figures, we compare the average auto-correlation relative errors between surrogate simulations and the DNS-based, as well as the average relative discrepancies between pairs of DNS-based simulations generated with different random seeds. Note that we use the term \textit{discrepancy} rather than \textit{error} because, even though the quantitative metrics are the same, the discrepancy is taken between pairs of DNS-based simulations only. Comparing the surrogate relative errors with the discrepancies allows assessing whether surrogate simulations differ significantly from DNS-based simulations generated under different stochastic conditions, given the chaotic nature of the underlying physics.

For $c_A = 0.2$, the surrogate achieves relative errors below 5\% during the training regime and up to 15-20\% in the extrapolation regime. Notably, the average relative error remains below the average discrepancy throughout both regimes, indicating that the surrogate simulations are virtually indistinguishable from solver-generated simulations within the extrapolation time horizon considered in this study. In fact, the surrogate appears slightly sub-chaotic, meaning that its predictions are statistically closer to the corresponding DNS-based than another DNS-based simulation (with a different random seed) would be. For $c_A = 0.3$, relative errors are slightly higher, reaching 20-30\% by the end of the extrapolation regime. After approximately 12$\mu$s, the relative error begins to exceed the average discrepancy, suggesting that the surrogate simulation deviates too much to be considered statistically equivalent to solver-generated simulation. This aligns with observations in Figure \ref{fig:qoi_030}, where certain QoIs, such as mass, interface perimeter, and mean ligament height, start to diverge from the DNS-based around the same time. Similarly, for $c_A = 0.4$, the relative auto-correlation error exceeds the average discrepancy upon entering the extrapolation regime, although it typically remains within 15-20\%.

As the surrogate rolls out auto-regressively in time, errors tend to accumulate. While these errors are relatively contained within the extrapolation time horizon considered in this paper, completely preventing error build-up remains challenging. This is primarily because our surrogate, like most neural operator models, is analogous to explicit numerical solvers (i.e. the field at a later time step depends, and only depends on the previous time step or a set of previous time steps) and thus suffers from similar pitfalls \cite{kovachki2021neural}. Besides, the surrogate is essentially data-driven, rather than physics-informed~\cite{Karniadakis:2021} (except for the hard-enforced boundary conditions). Moreover, there is no guarantee that the training data fully captures all physical patterns and dynamics that may emerge during the de-alloying process later on (i.e, in the extrapolation regime). This limitation likely explains why, for $c_A = 0.4$, the surrogate fails to capture the slowing down of the de-alloying process.

Assessing the reliability of the surrogate in the extrapolation regime is an open and challenging question. Visual inspection of the predicted fields can help identify blatantly unphysical artifacts, such as numerical instabilities. However, even visually plausible fields may not necessarily be physically accurate. Comparing surrogate predictions to solver-generated DNS-based solutions, which is what we do in this paper, is a straightforward validation strategy. It does become impractical for large-scale simulations, though. For instance, generating a single DNS-based simulation for $c_A = 0.4$ requires approximately three weeks on multiple CPU cores, making this approach increasingly unreasonable, and we are reaching its limit here. For surrogate simulations with a later time horizon than what is tested here, alternative validation methods, such as comparing certain QoIs to analytical tools or experimental data, could provide additional insights. These approaches are left for future work.

\FloatBarrier

\subsection{Model Setting Comparison}
\label{sec:comp}
\begin{figure}[!ht]
\centering
    \includegraphics[width=1.\textwidth]{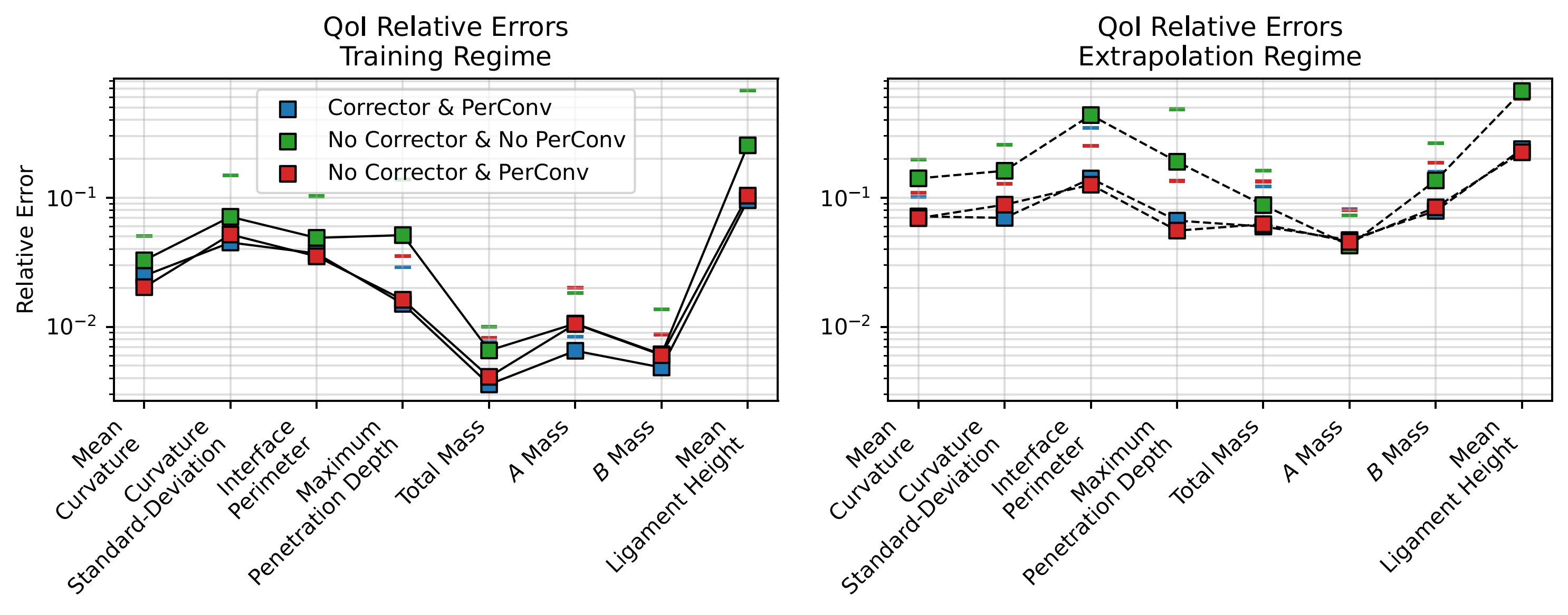}
    \caption{Ablation Study of the U-Net performances. The relative errors are averaged across all the concentrations available in the test set. \textit{Corrector} refers to the flood field corrector being applied in between each forward pass into the U-Net, as described in section \ref{flood_fill} \textit{PerConv} refers to the use of periodic convolutions (as described in section \ref{padding}). If no periodic convolutions are employed, the U-Net uses zero-padding instead. The horizontal bars indicate the maximum relative error for each setting across all 5 concentrations considered.}
    \label{model_comp}
\end{figure}

We now evaluate the effects of some of the components of the surrogate. We test the accuracy of the U-Net with and without periodic convolutions and flood-fill correction. The comparisons are summarized in Figure \ref{model_comp}, where relative errors are averaged across all concentrations available in the test set ($c_A = 0.2, 0.25, 0.30, 0.35,$ and $0.4$), with a breakdown provided for each QoI and for both the training and extrapolation regimes. The U-Net incorporating both periodic convolutions and the flood-fill corrector (i.e., the version used for all results in the previous section) generally achieves the highest accuracy across most QoIs. The flood-fill corrector slightly improves the accuracy of the QoIs in the training regime, particularly for the $A$ and $B$ mass. This improvement is expected, as the corrector primarily ensures that the $A$ and $B$ species fields remain rigorously accurate in regions that have not yet undergone corrosion. In the extrapolation regime, however, the performance of the model with and without the flood-fill correction is comparable. This is likely because as the de-alloying unfolds, the regions of the field still untouched by the corrosion (and primarily targeted by the corrector) become proportionally marginal, compared to the other regions. Removing periodic convolutions significantly deteriorates the results, especially in the extrapolation regime. This degradation is likely due to mismatched left and right boundaries, which introduce instabilities at the boundaries and potentially lead to unphysical artifacts in the QoI computations.

\begin{figure}[!ht]
\centering
    \includegraphics[width=1.\textwidth]{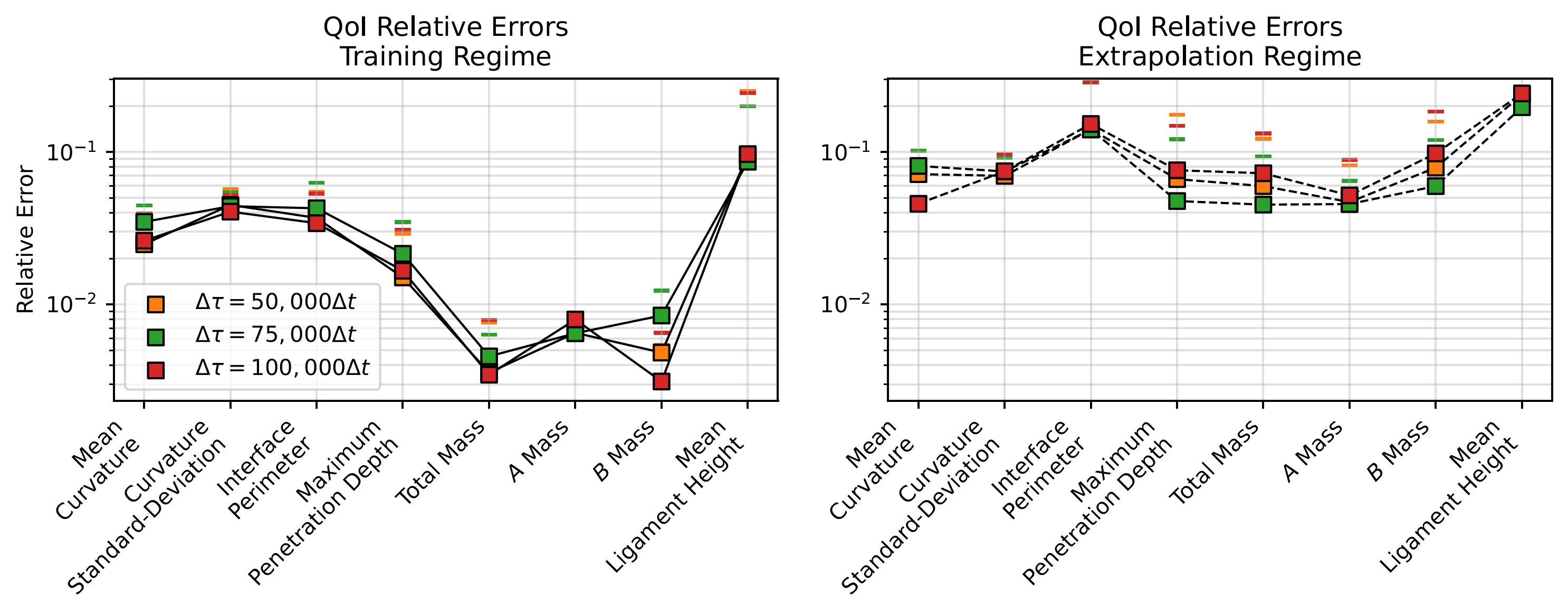}
    \caption{QoI relative errors for different skip step size $\Delta\tau$. One of the conditioning input parameter of the U-Net architecture the step size. This allows to chose how much time to skip between the input and the output of the model. During training, the U-Net learns to skip either 50,000 or 100,000 time steps, but can be tested on any arbitrary step size in between (or outside) this interval, for example 75,000 steps. The horizontal bars indicate the maximum relative error for each setting across all 5 concentrations considered.}
    \label{skip_comp}
\end{figure}

Figure \ref{skip_comp} shows a comparison of QoI relative errors given various time step skip sizes between the U-Net input and output. During training, the model is conditioned to skip either 50,000 or 100,000 time steps at a time, depending on the conditional parameter $\Delta\tau$. This allows us to evaluate whether skipping fewer or more time steps at-a-time affects the model's accuracy. Additionally, we test an intermediate step size of 75,000 time steps to assess the model's ability to interpolate between the step sizes seen during training. The U-Net performs similarly for both 50,000 and 100,000 time steps, particularly in the training regime. In the extrapolation regime, skipping 100,000 steps results in slightly higher errors compared to 50,000 steps, although the drop in accuracy is not substantial. The intermediate step size of 75,000 time steps yields slightly worse performance in the training regime, which is expected since the model was not explicitly trained to skip this particular number of steps and instead interpolates between the learned behaviors for smaller and larger step sizes. However, in the extrapolation regime, skipping 75,000 steps produces slightly better results than both 50,000 and 100,000 skips. This may be attributed to the inherent trade-off in auto-regressive models: skipping too many steps at-a-time can be challenging for the model to learn accurately, leading to higher errors, while skipping too few steps requires more forward passes to reach the same time horizon, potentially accumulating larger and larger errors. Thus, 75,000 steps may represent a balance between these extremes, serving as a \textit{sweet spot} for optimal performance in the extrapolation regime. Note that, although the surrogate skips through time in a discrete fashion (i.e. there is a gap of several thousand time steps between the input and the output), since the model can be conditioned by a continuous $\Delta\tau$ input, we can effectively predict the phase-field physics at any desired time. For example, if after running a surrogate simulation with a 50,000 time step increment, we now need to retrieve intermediate times, this can be easily done by using the already-computed time steps as a starting point for a U-Net conditioned on a carefully selected $\Delta\tau$.

\subsection{Surrogate Performance with Generative Initial Conditions}

\begin{figure}[!ht]
\centering
    \includegraphics[width=1.\textwidth]{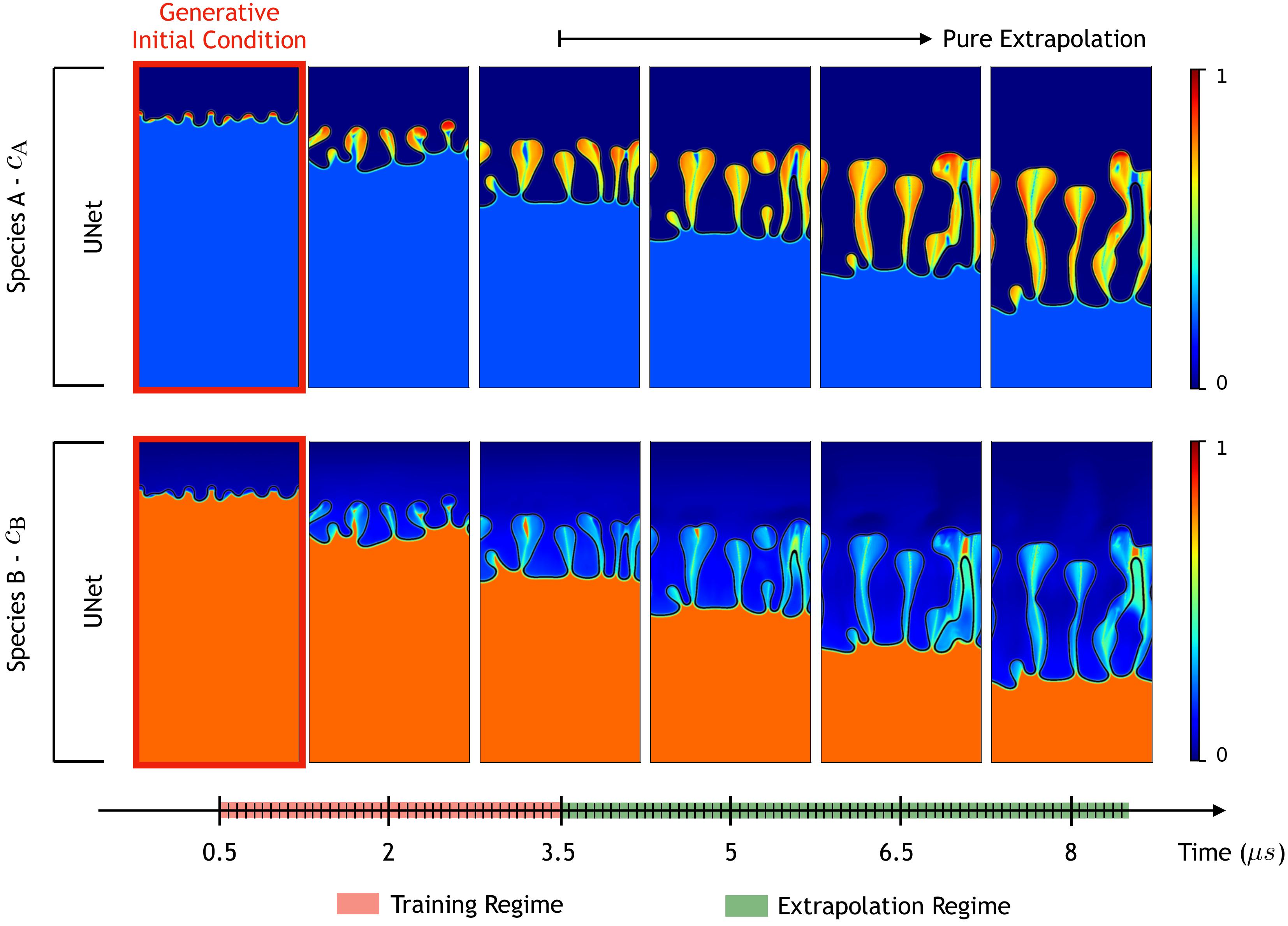}
    \caption{Concentration fields for species A and B. The ideal (pre-dealloying) concentration for A is $\pmb{c_A=0.2}$. The fields are obtained by sampling the diffusion model for generating the initial condition, and feeding it as initial input to the U-Net auto-regressive roll-out.}
    \label{field_genic}
\end{figure}

\begin{figure}[!ht]
\centering
    \includegraphics[width=1.\textwidth]{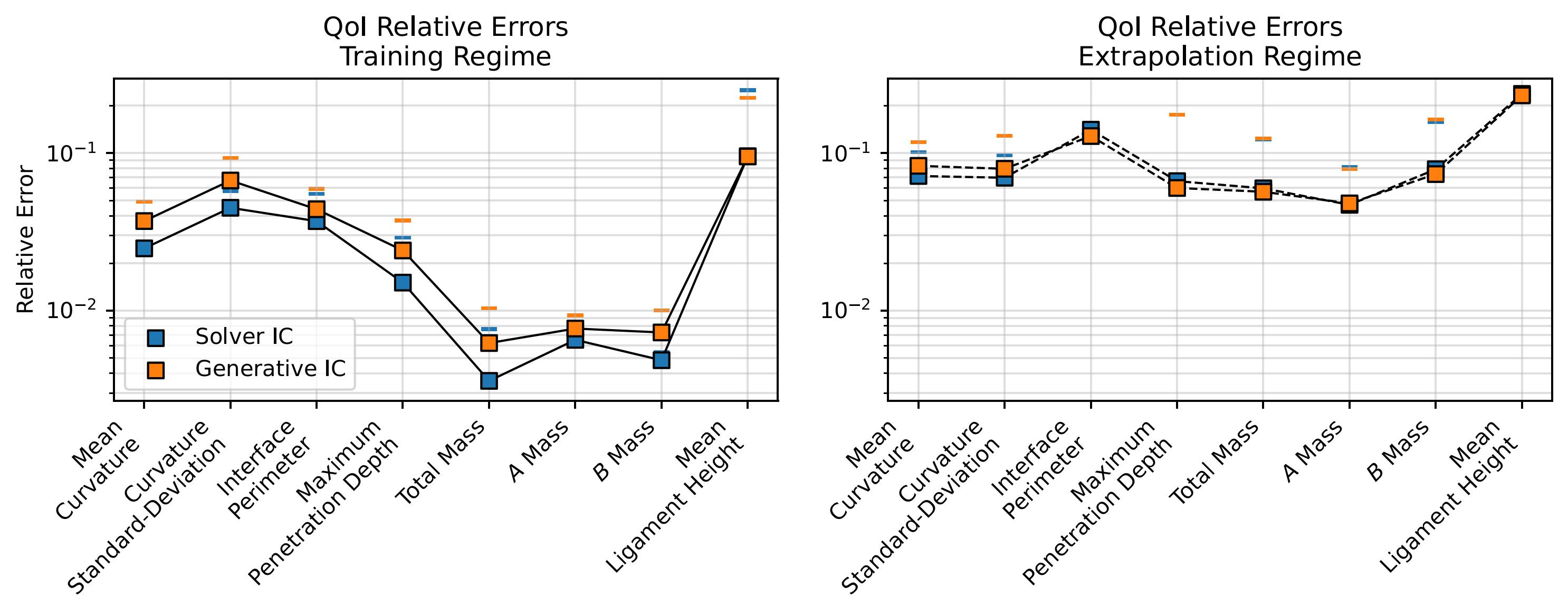}
    \caption{QoI relative errors of the U-Net surrogate, obtained based on both the generative (diffusion model-generated) initialization conditions, and the solver-generated (true) initial conditions. The horizontal bars indicate the maximum relative error for each setting across all 5 concentrations considered.}
    \label{qoi_errors_gen}
\end{figure}

\begin{figure}[!ht]
\centering
    \includegraphics[width=1.\textwidth]{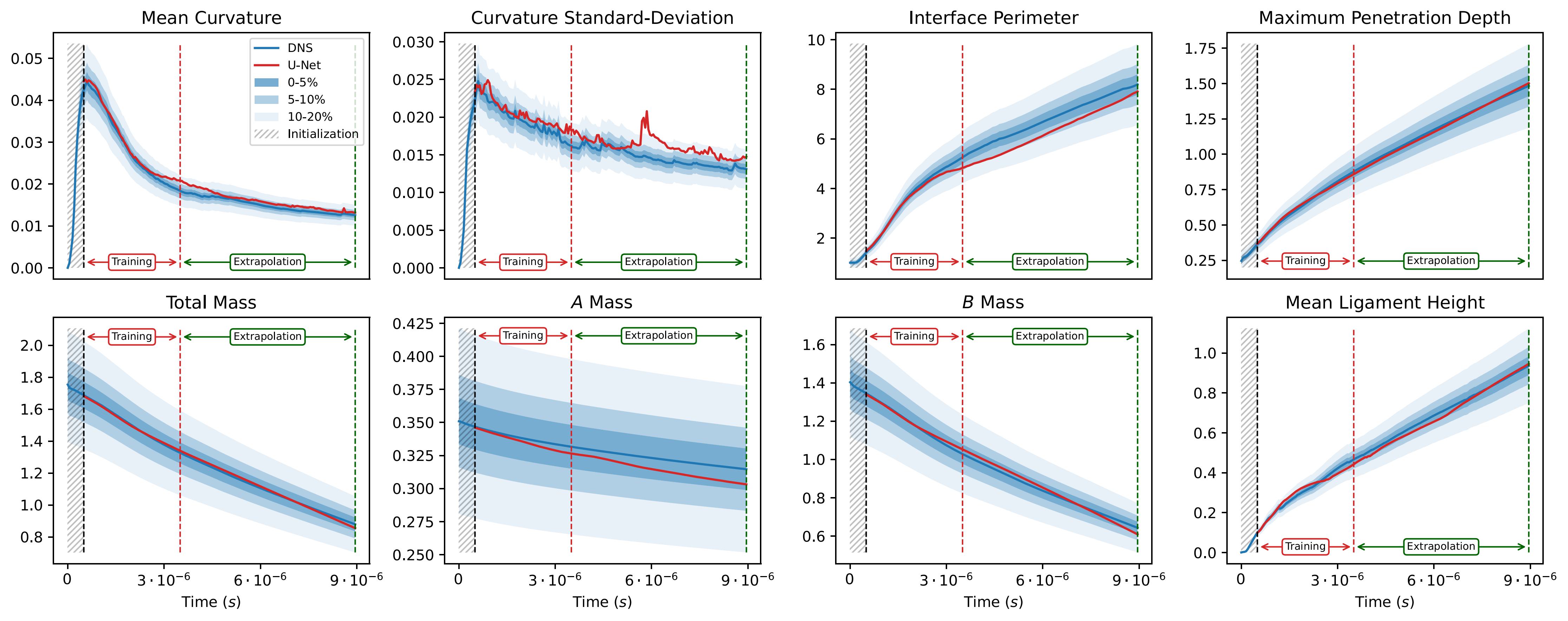}
    \caption{Mean QoI predictions for both the U-Net surrogate with generative initialization, in red (\raisebox{1.5pt}{\protect\tikz{\protect\draw[default line, very thick, line cap=round, Color2] (0, 0.) -- (0.33, 0.);}}) and the numerical solver, in blue (\raisebox{1.5pt}{\protect\tikz{\protect\draw[default line, very thick, line cap=round, Color1] (0, 0.) -- (0.33, 0.);}}). The ideal species is $\pmb{c_A=0.2}$}
    \label{qoi_020_gen}
\end{figure}

We now investigate the effect of using generative initial conditions in the surrogate simulations. Figure \ref{field_genic} showcases a surrogate simulation initialized with an initial condition sampled from the diffusion model for $c_A = 0.2$. Qualitatively, the general dynamics appear to be well respected, and comparable to both the DNS-based and the solver-initialized surrogate simulation shown in Figure \ref{fig:field_020}. Despite being synthetic, the generative initial condition appears physically realistic and does not negatively impact the subsequent evolution of the simulation. Figure~\ref{qoi_errors_gen} presents the QoI relative errors, averaged across all tested concentrations ($c_A = 0.2, 0.25, 0.3, 0.35,$ and $0.40$), for both diffusion model and solver-generated initial conditions. Using generative initialization slightly deteriorates the QoI relative errors in the training regime, but only by 1–2\% at most. In the extrapolation regime, the difference in accuracy between the two becomes essentially negligible. This behavior is likely due to the fact that generative initial conditions introduce minor errors at the beginning of the simulation, but the U-Net eventually converges to the same manifold of surrogate simulations as when employing (true) solver-generated initial conditions. Figure \ref{qoi_020_gen} presents the mean predictions for each QoI, along with the relative error range for $c_A = 0.2$, and using diffusion model-generated initial conditions. The predictions are almost indistinguishable from those obtained using solver-generated initial conditions (as shown in Figure~\ref{fig:qoi_020}), confirming that synthetic initial conditions do not affect the surrogate's performance. This finding means that, in practice, time-consuming solver-generated initialization (as done in previous work~\cite{bonneville_2025, Oommen:2024}) can be entirely replaced by generative model-based initialization, enabling significant time savings without compromising accuracy.

 \FloatBarrier
 
\subsection{Acceleration Performance}

We now evaluate the computational time savings achieved with the surrogate model, as this is the primary motivation for developing such models. A single forward pass through the U-Net takes approximately 0.15 seconds on a single A100 GPU. This computational cost is constant regardless of the time increment being skipped between the input and output (i.e., regardless of $\Delta\tau$). Note that this time includes the processing required for the flood-fill corrector. In contrast, the numerical solver requires an average of 59 minutes on 64 CPU cores to compute 50,000 time steps. Consequently, for a U-Net configured to skip 50,000 steps per forward pass, the speed-up is approximately 23,700$\times$, and this increases to 47,300$\times$ when skipping 100,000 steps.

Generating initial conditions with the numerical solver is also computationally expensive. For initialization at $t = 0.5 \, \mu$s, the solver requires approximately 9 hours and 51 minutes, and this time doubles for initialization at $t = 1 \, \mu$s. Using the diffusion model however, generating one initial condition takes only 71 seconds. Note that this time can be further reduced by batching multiple initial conditions, leveraging the parallel processing capabilities of GPUs. For instance, generating 20 initial conditions in a single batch takes approximately 258 seconds, or 13 seconds per sample (nearly 6$\times$ faster than generating them individually).

To illustrate the overall time savings, consider the LMD simulation shown in figure \ref{fig:field_020}, with $c_A = 0.2$ and run for 9$\mu$s of real time (equivalent to 9,000,000 time steps). Using the numerical solver, this simulation takes approximately 177.5 hours, or 7.5 days, to complete. With the surrogate model augmented by the diffusion model and configured to skip 50,000 time steps per forward pass, the same simulation takes only 38.5 seconds, resulting in a speed-up of 16,600$\times$. If the surrogate is configured to skip 100,000 time steps per forward pass, the speed-up increases to 24,800$\times$. For a more demanding case like the simulation depicted in figure \ref{fig:field_040}, with $c_A = 0.4$ run for 24 microseconds, the numerical solver takes approximately 20 days to complete. With the surrogate model configured to skip 100,000 time steps per forward pass, the simulation takes only 47.5 seconds, yielding an effective speed-up of 35,900$\times$.

\section{Conclusion}

In this paper, we introduced a U-Net-based auto-regressive model specifically designed to extrapolate both in time and space for LMD simulations. By incorporating circular padding and correcting non-corroded regions after each forward pass, we efficiently enforce boundary conditions and maintain stability, enhancing the model's robustness during long-term simulations. Furthermore, by implementing all components of the model using convolutional layers and tuning only the channel dimensions rather than the spatial dimensions during conditioning, the model becomes independent of the input/output field size. This enables us to extend the spatial dimensions beyond those used during training, and facilitates significant extrapolation in space and ultimately time. The conditioning mechanism further enhances the model's flexibility, and here we specifically show that the model is able to capture dramatic changes in corrosion morphology and rate as a function of initial alloy composition. This methodology offers a pathway towards conditioning the model on its thermodynamic and kinetic inputs. Such inputs are known to control the initial stages of dealloying phase field simulations \cite{Lai:2022,KerrBieberdorf2024}, but their effects on large scale microstructure evolution have not been explored.

Our model demonstrates robust extrapolation capabilities, achieving high accuracy even when tested on time horizons approximately three times longer than those present in the training data. The model is particularly accurate for lower $c_A$ values, achieving relative errors below 10\% in the extrapolation regime. This accuracy tends to slightly deteriorate for higher $c_A$ concentrations, likely because the model may not have been exposed to all the possible physical patterns during training. Additionally, the model can be initialized using synthetic initial conditions generated by diffusion models, with minimal loss in accuracy, eliminating the need for computationally expensive solver-based initialization. Ultimately, the proposed model achieves substantial speed-ups, up to $10^4-10^5\times$ faster than traditional solvers, enabling simulations to reach time horizons that are otherwise infeasible within reasonable computational budgets. This also reduces the cost of generating training data, as training simulations can be run for shorter time spans, with the model extrapolating beyond those spans during inference.

The extent to which the current model can reliably extrapolate remains unclear. During auto-regressive inference, errors tend to accumulate, and under certain conditions (e.g. specific material properties, time horizons, etc.), the model's predictions in extrapolation may become highly inaccurate. This is in fact likely if there are physical patterns emerging at later times that differ significantly from those captured in the early training data. Understanding potential failure modes in extrapolation remains a challenge, especially for extreme extrapolation scenarios where direct numerical simulations are unavailable for comparison. Proper validation of such cases may require alternative techniques, such as analytical tools or experimental data, although this is left to future work.

\section*{Acknowledgments}

This work was supported by the U.S. Department of Energy, Office of Science, Office of Advanced Scientific Computing Research through the FASTMath Institute and jointly with the Office of Nuclear Energy through the SciDAC project on Simulation of the Response of Structural Metals in Molten Salt Environment.
This article has been co-authored by employees of National Technology and Engineering Solutions of Sandia, LLC under Contract No. DE-NA0003525 with the U.S. Department of Energy (DOE). The employees co-own right, title and interest in and to the article and are responsible for its contents. The United States Government retains and the publisher, by accepting the article for publication, acknowledges that the United States Government retains a non-exclusive, paid-up, irrevocable, world-wide license to publish or reproduce the published form of this article or allow others to do so, for United States Government purposes. The DOE will provide public access to these results of federally sponsored research in accordance with the DOE Public Access Plan \url{https://www.energy.gov/downloads/doe-public-access-plan}. Sandia Release Number: SAND2025-14371O.
Los Alamos National Laboratory, United States, an affirmative action/equal opportunity employer, is operated by Triad National Security, LLC, for the National Nuclear Security Administration of the U.S. Department of Energy under Contract No. 89233218CNA000001.
Lawrence Berkeley National Laboratory is supported by the DOE Office of Science under contract no. DE-AC02-05CH11231. This study made use of computational resources of the National Energy Research Scientific Computing Center (NERSC), which is also supported by the Office of Basic Energy Sciences of the US Department of Energy under the same contract number.

\bibliographystyle{unsrt}  
\bibliography{references}  

\end{document}